\documentclass[]{aa}
\usepackage{graphicx}
\usepackage{txfonts}

\usepackage{natbib}
\bibpunct{(}{)}{;}{a}{}{,}
\begin{document}
   \title{Fluctuations of the intergalactic UV background towards two lines of sight\thanks{Based on observations made with the VLT/Kueyen telescope ESO, Paranal, Chile}}

   \subtitle{}

   \titlerunning{Fluctuations of the intergalactic UV background}
   \author{C. Fechner
          \inst{1}
          \and
          D. Reimers\inst{1}
          }

   \offprints{C. Fechner}

   \institute{Hamburger Sternwarte, Universit\"at Hamburg,
              Gojenbergsweg 112, 21029 Hamburg, Germany\\
              \email{[cfechner,dreimers]@hs.uni-hamburg.de}
             }

   \date{Received ; accepted}

   \abstract{We present a reanalysis of the \ion{He}{ii} Ly$\alpha$ absorption towards the quasars \object{HS~1700+6416} and \object{HE~2347-4342} using new high $S/N$, optical observations.
An alternative analysis method is applied, which fits the high quality, optical \ion{H}{i} data directly to the \ion{He}{ii} spectrum.
The results are compared to those inferred from standard line profile analyses.
This new method enables us to derive redshift scales characterizing the fluctuations of the column density ratio $\eta$.
We find $\eta$ changing smoothly with redshift on typical scales of $\Delta z \sim 0.01 - 0.03$ corresponding to $8 - 24\,h^{-1}\,\mathrm{Mpc}$ comoving.
The real length scales of variations of the column density ratio might
be even larger, since part of the fluctuations may be caused by noise
in the \ion{He}{ii} data and by systematic effects due to the applied method. 
However, variations on small scales cannot be ruled out completely.
For $\eta$ variations on scales of a few Mpc an amplitude of about $\pm 1.5\,\mathrm{dex}$ cannot be excluded.
The data shows an apparent correlation between low $\eta$ regions and the presence of metal line absorbers, which corresponds to the more general correlation of low $\eta$ and strong \ion{H}{i} absorption.
Thermal line broadening is suggested as a probable explanation for this apparent correlation, since both fit methods would severely underestimate $\eta$ for absorbers with $N_{\ion{H}{i}} \gtrsim 10^{13}\,\mathrm{cm}^{-2}$ if the line width was dominated by thermal broadening.
Indeed, lines located close to the cut-off of the $b(N)$ distribution yield lower column density ratios compared to the whole sample, in particular if high density absorbers are considered.
We argue that the apparent correlation of $\eta$ with the strength of the \ion{H}{i} absorption is caused by insufficient consideration of thermal broadened lines by the standard analysis.
As unbiased value of the column density ratio, we find $\eta \sim 80$ in agreement with previous estimates.

   }

   \maketitle
%

\section{Introduction}

The metagalactic ionizing radiation plays an important role for QSO absorption line studies probing the intergalactic medium (IGM).
The IGM is highly photoionized and the UV background governs its ionization state.
Estimates of the intergalactic metallicity are highly dependent on the presumed ionizing radiation \citep[e.g.][]{aguirreetal2004}.
Nevertheless, the shape of the UV background is inaccessible to direct observation.
Theoretical models based on the number, distribution, and spectral properties of  underlying sources like AGN and starburst galaxies have been computed by e.g.\ \citet{haardtmadau1996, haardtmadau2001, fardaletal1998, bianchietal2001}.
Efforts to derive the shape of the ionizing radiation from the observed column densities of different metal ions via photoionization calculations have been made recently by \citet{agafonovaetal2005}.
The observational studies suggest deviations from the shape predicted by ``classical'' theoretical models \citep[see also][]{boksenbergetal2003, fechneretal2006a}.
Additionally, recent theoretical work \citep{schaye2004, miraldaescude2005} pointed out that local radiation sources may be important for high density absorbers.

Apart from column density ratios of metal line absorption, the Ly$\alpha$ transition of \ion{He}{ii} ($\lambda_0 = 303.7822\,\mathrm{\AA}$) provides a further possibility to probe the UV background observationally.
The column density ratio $\eta = N_{\ion{He}{ii}}/N_{\ion{H}{i}}$ depends on the photoionization rates of \ion{H}{i} and \ion{He}{ii} and thus on the ionizing radiation at the \ion{H}{i} (1\,Ryd) and \ion{He}{ii} (4\,Ryd) ionization threshold \citep[e.g.][]{fardaletal1998}.
Furthermore, \ion{He}{ii} probes the low density phase of the IGM, since \ion{He}{ii} is more difficult to ionize than \ion{H}{i} due to lower fluxes and cross sections at its ionizing threshold and the faster recombination of \ion{He}{iii} compared to \ion{H}{ii}.

Direct observations of \ion{He}{ii} Ly$\alpha$ absorption are very difficult due to the limited number of unabsorbed quasars \citep{picardjakobsen1993}.
Up to now, six objects have been observed, showing \ion{He}{ii} absorption in their UV spectra.
Only two of them (HE~2347-4243 at $z = 2.885$ and HS~1700+6416 at $z = 2.72$) are bright enough for high resolution spectroscopy with FUSE \citep{krissetal2001, shulletal2004, zhengetal2004, reimersetal_fuse, fechneretal2006b}.
The results from these data suggested that the UV background might fluctuate on small scales ($\Delta z \sim 0.001$) with an amplitude of several orders of magnitude ($\eta \lesssim 10$ up to $\gtrsim 1000$).
Probable reasons for these fluctuations are the properties of the QSOs like the spread in the spectral indices \citep[e.g.][]{telferetal2002} and finite lifetimes, local density variations of the IGM \citep{miraldaescudeetal2000} or filtering of radiation by radiative transfer effects \citep{maselliferrara2005}.
Furthermore, it has been shown that also the noise level of the observed spectra contributes to the scatter in $\eta$ \citep{fechneretal2006b, liuetal2006}.

Due to the limited quality of the FUSE data ($S/N \sim 5$) it is important to examine in detail the applicability of the adopted analysis methods.
\citet{shulletal2004} applied an apparent optical depth method, while all other studies are based on line profile fitting \citep{krissetal2001, zhengetal2004, reimersetal_fuse}.
\citet{foxetal2005} showed that the apparent optical depth method overestimates the column density in case of low signal-to-noise.
Furthermore, it is only valid for unsaturated lines \citep{savagesembach1991}. 
The applicability of line profile fitting was investigated by \citet{boltonetal2006}, who find that it leads to reliable results (but see discussion below), and the median provides more confident results than the mean.
According to their study, the fluctuations in $\eta$ are due to the small number of quasars that contribute to the \ion{He}{ii} ionization rate at any given point.
They also conclude that a small fraction of space have to be exposed to a harder radiation, since they failed to reproduce column density ratios $\lesssim 1$, which are measured by \citet{zhengetal2004}.
A similar finding is reported by \citet{liuetal2006}.

Here, we re-analyze the \ion{He}{ii} Ly$\alpha$ forest towards HE~2347-4342 and HS~1700+6416 applying a new method, which fits the optical data directly to the FUSE spectra.
This approach minimizes effects of noise mimicking $\eta$ variations when applying the apparent optical depth method, and it avoids the subjectivity of the line profile modeling, where the number of assumed components are due to arbitrary choice.
In order to provide an alternative estimate of the scales on which $\eta$ fluctuates, we follow several approaches.
Furthermore, we investigate the effect of line broadening on the inferred $\eta$ values.

The paper is organized as follows:
New high quality, optical data of the \ion{H}{i} Ly$\alpha$ forest towards HE~2347-4342 are presented in Sect.\ \ref{observations3}.
In Sect.\ \ref{method}, we introduce the fit procedure (called spectrum fitting), which is applied to the data in Sect.\ \ref{results2}.
The results, possible effects due to thermal line broadening, and their implications are discussed in Sect.\ \ref{discussion3}.
We conclude in Sect.\ \ref{conclusions3}.

\section{Data}\label{observations3}

HE~2347-4342 was observed with UVES at the second VLT unit telescope (Kueyen) during 13 nights at two epochs in 2000 and 2003.
The details of the settings are presented in Table \ref{he2347_obs}.
In 2000, the observations were carried out as a part of the Large Program "The Cosmic Evolution of the Intergalactic Medium" (116.A-0106A) in the UVES dichroic modes resulting in a total wavelength coverage of $\sim 3000 - 10\,400\,\mathrm{\AA}$.
Additional observations were performed in 2003 in the blue with a central wavelength of $4370\,\mathrm{\AA}$.
The slit width of $1\arcsec$ and binning $2\times 2$ were chosen to match the setting of the Large Program data.
The resolution is $R \approx 45\,000$.

\begin{table}[!ht]
  \begin{center}
    \caption{Observation diary of HE~2347-4342. All exposures use a $1\arcsec$ slit and
 $2\times 2$ binning. The seeing conditions were very good with $0.5\arcsec - 1.0\arcsec$.
}
    \vspace{0.2cm}
    \begin{tabular}{c l l r}
      \hline
      \hline
       date & mode & $\lambda_{\mathrm{cen}}$ (nm) & $t_{\mathrm{exp}}$ (s) \\
      \hline
      2000/10/07 & DICHR\#1 & 346/580 & 14\,400 \\
      2000/10/08 & DICHR\#1 & 346/580 &  7\,200 \\
      2000/10/08 & DICHR\#2 & 437/860 &  7\,200 \\
      2000/10/09 & DICHR\#2 & 437/860 &  7\,200 \\
      2000/11/19 & DICHR\#2 & 437/860 &  7\,200 \\
      2003/06/05 & BLUE & 437 &  7\,200 \\
      2003/06/06 & BLUE & 437 & 10\,800 \\
      2003/06/07 & BLUE & 437 &  6\,900 \\
      2003/06/08 & BLUE & 437 & 10\,300 \\
      2003/06/12 & BLUE & 437 &  3\,130 \\
      2003/06/17 & BLUE & 437 &  6\,600 \\
      2003/06/24 & BLUE & 437 &  6\,600 \\
      2003/07/06 & BLUE & 437 &  3\,700 \\
      2003/08/05 & BLUE & 437 &  3\,244 \\
      \hline
    \end{tabular}
    \label{he2347_obs}
  \end{center}
\end{table}

The data reduction was performed at Quality Control Garching using the UVES pipeline Data Reduction Software \citep{ballesteretal2000}.
The vacuum-barycentric corrected spectra were co-added.
Thus, for the wavelength range $3800 - 4950\,\mathrm{\AA}$ the total exposure time is 28.24\,hrs resulting in a signal-to-noise ratio of $S/N \approx 100$ in the Ly$\alpha$ forest.

The UV data of HE~2347-4342 data have been taken with the FUSE satellite in 2000.
The data were first presented by \citet{krissetal2001}, and was newly reduced and re-analyzed twice \citep{shulletal2004, zhengetal2004}.
In this study we use the spectrum of \citet{zhengetal2004}, and refer to their work for details of the data reduction.
The resulting spectrum covers the spectral range $900 - 1188\,\mathrm{\AA}$.
The signal-to-noise ratio is $S/N \approx 5$ on a bin size of $0.5\,\mathrm{\AA}$ in the best part ($1088-1134\,\mathrm{\AA}$) and slightly less elsewhere. 
The resolution is $R \approx 20\,000$.

A low resolution STIS spectrum was taken simultaneously with the FUSE observations in August and October 2000 using the gratings G140L and G230L.
The spectra cover the wavelength range $1120 - 3170\,\mathrm{\AA}$ and are used to extrapolate the continuum to the FUSE spectral range. 
We estimate the continuum considering galactic extinction, Lyman limit absorption, and the intrinsic spectral energy distribution of the QSO.
The extinction curve is adopted from \citet{cardellietal1989} assuming $E(B-V) = 0.014$ \citep{schlegeletal1998}.
In addition, there is a Lyman limit system (LLS) on the line of sight at $z \approx 2.735$.
From the Lyman break we estimate the optical depth $\tau \sim 1.8$ leading to $\log N_{\mathrm{LLS}} \approx 17.46$ following \citet{mollerjakobsen1990}, which is in good agreement with the results of \citet{reimersetal1997}.
However, the contributed optical depth of the LLS to the absorption in the FUSE spectral range is negligible ($\tau < 0.08$ at $\lambda < 1200\,\mathrm{\AA}$) thanks to the $\nu^{-3}$ dependence of the Lyman continuum opacity. 
Considering the LLS optical depth and the extinction law, we estimate the intrinsic QSO power law to have a spectral slope of $\alpha \approx -0.46$. 

The UV spectrum of HS~1700+6416 taken with FUSE was first presented in \citet{reimersetal_fuse}. We use the reduced and normalized data set from \citet{fechneretal2006b}, which has a resolution of $R \approx  20\,000$ and a signal-to-noise ratio of $\sim 5$.
The corresponding \ion{H}{i} Ly$\alpha$ forest was observed with Keck/HIRES.
For a detailed presentation of the optical data we refer to \citet{fechneretal2006a, fechneretal2006b}.
The resolution is $R \approx 38\,000$ and $S/N \approx 100$.

\section{Method}\label{method}

Previous studies of the \ion{He}{ii} Ly$\alpha$ forest are based on a profile-fitting procedure \citep{krissetal2001, zhengetal2004, reimersetal_fuse} or the apparent optical depth method \citep{shulletal2004}.
Both methods suffer from different shortcomings.
The apparent optical depth method leads to reliable results only if lines are optically thin \citep{savagesembach1991}.
In case of the \ion{He}{ii} forest many profiles are saturated.
Thus, the validity of the apparent optical depth method is limited.
Furthermore, the results strongly depend on the noise, which may mimic fluctuations of $\eta$.

In case of fitting line profiles to the data the \ion{H}{i} spectrum is modeled by Doppler profiles.
The derived line parameters are used to fit the \ion{He}{ii} data as well.
In doing so, the line widths have been assumed to be dominated by turbulent broadening, i.e.\ $b_{\ion{He}{ii}} = b_{\ion{H}{i}}$.
This assumption is also made implicitly in the apparent optical depth method, and is inherent in our spectrum fitting method as well.
However, fitting line profiles need the subjective decision how many components are used to model the observed features.
Particularly, this problem gets significant for weak lines and complicated blends, where components may get lost or the decomposition is insufficient.
Problems in the \ion{H}{i} model may cause a poor description of the corresponding \ion{He}{ii} absorption, where the features are stronger but the data quality is worse. 
Therefore, apparent variations in $\eta$ may be artifacts of an inadequate decomposition.

With the purpose to avoid a strong dependence on the noise as well as the subjectivity of a model, we develop a new method to estimate $\eta$.
It is based on the idea to compare the \ion{He}{ii} and the \ion{H}{i} Ly$\alpha$ forest directly without the need to provide a model of the data and without using a pixel by pixel (or bin by bin) technique which would be affected by the noise level of the data.
Therefore, we fit the whole \ion{H}{i} spectrum to the \ion{He}{ii} data by scaling it by $\eta = 4\cdot\tau_{\ion{He}{ii}}/\tau_{\ion{H}{i}}$ \citep{miraldaescude1993}, which is then the only free parameter.

Since an unbiased \ion{H}{i} spectrum is needed, we identify the metal line systems in the optical data, fit Doppler profiles to the spectral range of interest and subtract them from the spectrum.
Furthermore, we confine ourselves to the pure Ly$\alpha$ forest, and only the data between Ly$\alpha$ and Ly$\beta$ emission are suitable.
In case of HE~2347-4342 these are wavelengths $\gtrsim 4000\,\mathrm{\AA}$ in the optical and $\gtrsim 1000\,\mathrm{\AA}$ in the UV.
Strong \ion{O}{vi} absorption arising from a complex system associated with the QSO is present shortward of $\sim 4050\,\mathrm{\AA}$.
Due to the complexity of the system and severe blending with forest lines \citep[for a more detailed discussion see][]{fechneretal2004}, we restrict the redshift range to $2.33 < z < 2.91$.

In preparation of a fit, the wavelength scales of the datasets are aligned scaling the optical data by a factor of $1215.6701\,\mathrm{\AA}/303.7822\,\mathrm{\AA} \approx 4.00178$.
Furthermore, the resolution of the optical spectrum is changed to match the lower resolution of the FUSE data by a convolution with a Gaussian.
The method assumes implicitly that line broadening is dominated by turbulent motion and the line widths are the same for \ion{H}{i} and \ion{He}{ii}.
Since the operation of convolution is commutative, the width of the Gaussian has to be $\sigma = (\sigma_{R=20\,000}^2 - \sigma_{R=45\,000}^2)^{1/2}$.

In order to perform the actual fit, a wavelength portion is chosen and $\log\eta$ is estimated to minimize
\begin{equation}\label{chi2}
  \chi^2 = \frac{1}{n}\sum_{i=1}^{n}\frac{\left(f_{\ion{He}{ii},\,i} - \exp\left(-\frac{\eta}{4}\cdot\tau_{\ion{H}{i},\,i}\right)\right)^2}{\sigma_{\ion{He}{ii},\,i}^2 + \sigma_{\ion{H}{i},\,i}^2}\,\mbox{,}
\end{equation}
where $f_{\ion{He}{ii}}$ and $\tau_{\ion{H}{i}} = -\ln f_{\ion{H}{i}}$ are the normalized observed flux and the optical depth of \ion{He}{ii} and \ion{H}{i}, respectively, with the uncertainties $\sigma_{\ion{He}{ii}}$ and $\sigma_{\ion{H}{i}}$; $n$ is the number of pixels in the chosen wavelength interval.
The $1\,\sigma$ uncertainty of the fitted $\log\eta$ is estimated finding the values at which $\chi^2 = \chi^2_{\mathrm{\min}} \pm \Delta\chi^2$ with $\Delta\chi^2 = 1.0$.
We restrict the parameter space to values $0.0 \le \log\eta \le 4.5$.

\begin{figure*}
  \centering
  \resizebox{\hsize}{!}{\includegraphics[bb=40 35 473 780,clip=,angle=-90]{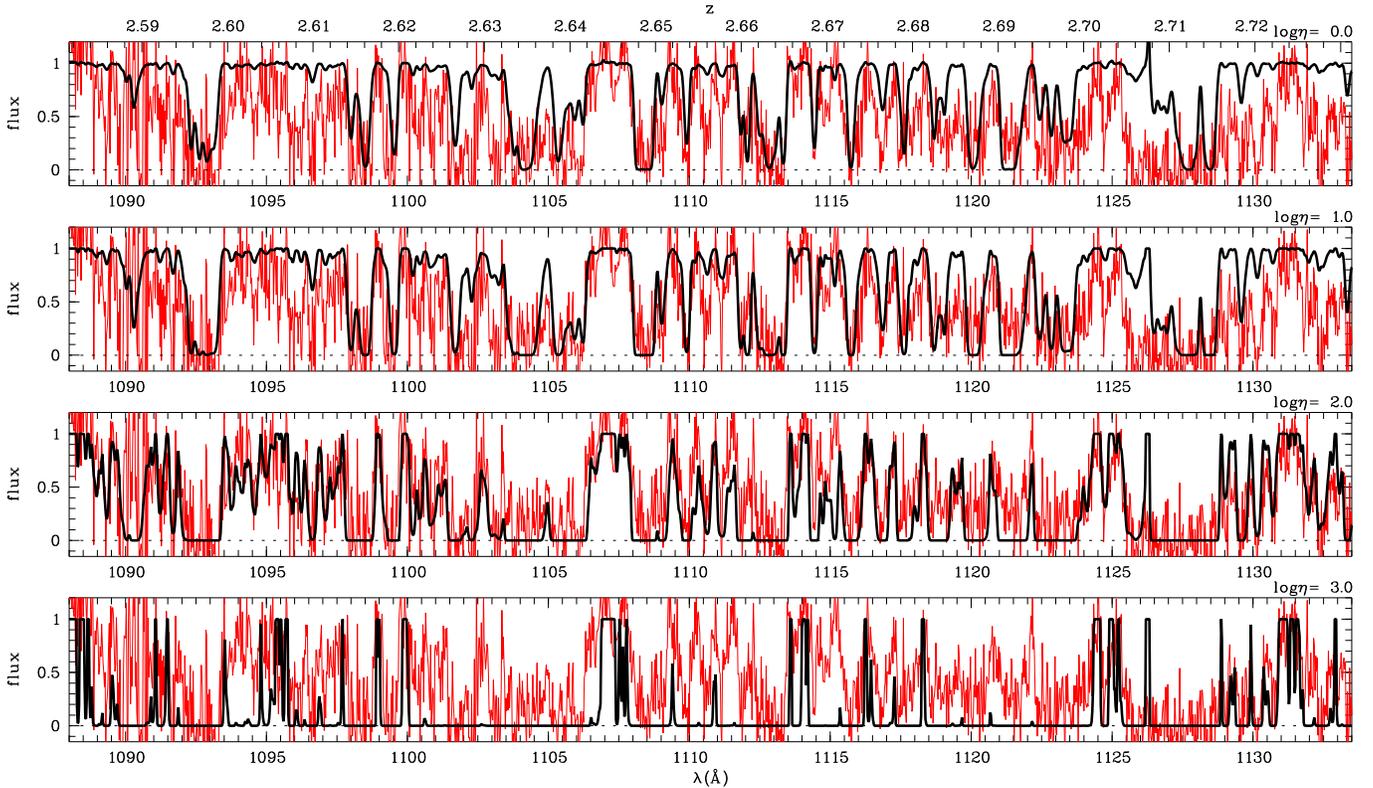}}
  \caption{Comparison of the observed \ion{He}{ii} data towards HE~2347-4342 to the metal subtracted, convolved \ion{H}{i} spectrum scaled by $\log\eta = 0.0,\, 1.0,\, 2.0,$ and $3.0$.
Apparently, there are regions modeled well by $\log\eta \approx 1.0$ (e.g.\ $1112 - 1117\,\mathrm{\AA}$) or $\log\eta \approx 2.0$ (e.g.\ $1096 - 1103\,\mathrm{\AA}$ and $\sim 1125\,\mathrm{\AA}$), respectively, while $\log\eta \approx 3.0$ would lead only to sparse regions with non-zero UV flux, which is inconsistent with the observations in this redshift range.
Features that appear nearly overestimated by $\log\eta = 0.0$ (at $\sim 1104\,\mathrm{\AA}, 1108.5\,\mathrm{\AA}$, and $1121.5\,\mathrm{\AA}$) correspond to weak metal line systems (see text).
  }
  \label{example}
\end{figure*}

The idea of fitting the spectra directly is illustrated in Fig.\ \ref{example} for the spectral range $1088.0 - 1133.5\,\mathrm{\AA}$ of HE~2347-4342, where the data quality of the UV spectrum is best.
Each panel shows the normalized FUSE data in comparison to the aligned, metal line corrected \ion{H}{i} spectrum scaled by $\log\eta = 0.0$ (i.e.\ unscaled), $1.0$, $2.0$, and $3.0$, respectively. 
Obviously, some regions are consistent with $\log\eta \approx 1$ (e.g.\ $1112 - 1117\,\mathrm{\AA}$) or $\log\eta \approx 2$ (e.g.\ $1096 - 1103\,\mathrm{\AA}$).
Whereas scaling the \ion{H}{i} spectrum by $\log\eta \approx 3$ leads to almost complete absorption with few transmission windows, which is apparently not consistent with the \ion{He}{ii} observations in the presented wavelength range.
The overall fit of this spectral range leads to $\log\eta = 1.81^{+0.48}_{-0.50}$ with $\chi^2 = 3.34$.
However, the considered portion corresponds to $\Delta z \approx 0.15$, and obviously, $\eta$ varies on smaller scales. 

The definition of a scale in redshift, on which $\eta$ is supposed to appear constant, is the most crucial point.
The required length scale is given by the number of pixels $n$ that are considered.
If $n$ is chosen too small, the result is severely affected by the noise level of the FUSE data.
If $n$ gets very small, the procedure approaches the apparent optical depth method.
On the other hand, possible small-scale fluctuations will be smoothed and therefore remain undetected if the length scale is chosen too broad.

In order to investigate possible effects on the results, we follow four different approaches, (i) constant binning, (ii) binning ``per eye'', (iii) binning by $\chi^2$-estimation, and (iv) averaging around each pixel.
We introduce these procedures in the following.

(i) Most simply, bins of a fixed size are adopted.
In this case, the bin size (we use $0.5$ and $1.0\,\mathrm{\AA}$) and the positioning of the bins are free parameters.
Therefore, fluctuations are investigated only on the given scale.

(ii) In order to find a length scale, which is not fixed a priori, we define the fit intervals by eye.
Intervals are fitted successively starting at the low wavelength edge of the data.
The size of the intervals is chosen to be as large as possible to show no obvious fluctuations of $\eta$.
The advantage is that apparent changes of $\eta$ as illustrated by Fig.\ \ref{example} can be considered.
On the other hand, the choice of the size of the intervals is made very subjectively.

(iii) The more objective way is to start with a certain number of
pixels at the low wavelength limit of the spectrum and estimate the $\chi^2$ of the fit.
More pixels are added to the fit interval as long as the $\chi^2$ of the fit decreases.
If the $\chi^2$ increases again, the interval is defined and the next evaluation starts.
Test calculations show that in the case of HE~2347-4342 at least 15 pixels, corresponding to $0.375\,\mathrm{\AA}$ have to be used for the first evaluation to avoid confusion due to the noise.
This means the smallest detectable scales have a size of $\Delta z \sim 0.001$.

(iv) Alternatively, we average over an interval of a certain size centered on every pixel.
This procedure corresponds to a pixel by pixel estimation smoothed by averaging over a certain number of adjacent pixels.
However, the results again depend on the number of pixels used, i.e.\ the degree of smoothing.
A good choice appears to be 40 pixels corresponding to $1.0\,\mathrm{\AA}$.

Another crucial point is the continuum estimation of the optical data.
If the continuum is located too low, absorption of \ion{H}{i} will be removed and the fitted $\eta$ value will be overestimated, while $\eta$ will be underestimated if the continuum level is too high.
In order to determine an optimal continuum we perform the continuum fit simultaneously with a line profile fit using the line fitting program CANDALF \citep[also used e.g.\ by][]{fechneretal2006b}.
The fit program models the continuum by Legendre polynomials.
The simultaneous estimation of continuum and line parameters is supposed to lead to a more accurate continuum level than an a priori normalization.
Using the formal errors of the Legendre parameters a statistical uncertainty of the continuum can be estimated.
We find a statistical error of less than $\sim 1\,\%$.
Additional systematic uncertainties due to the considered number of components are hard to quantify.
Because of the exceptionally high signal-to-noise ratio of the optical data a realistic estimate of the systematic error should be $\lesssim 1\,\%$.

Test calculations show that a continuum error of $1\,\%$ leads to an uncertainty in $\eta$ of $\pm 0.1\,\mathrm{dex}$.
This value is small in comparison to the fit errors due to the noisy \ion{He}{ii} data which are on average $\pm 0.4\,\mathrm{dex}$ or even larger (see below).
Therefore, possible uncertainties introduced by the continuum fit will be neglected in the following.
However, the presented errors of $\log\eta$ might be about 3\,\% larger due to the uncertainties in the continuum estimation.
A more conservative estimate of the continuum fit error ($\sim 2\,\%$) would result in an additional uncertainty in $\eta$ of roughly $0.2\,\mathrm{dex}$ which would enlarge the given errors by about 12\,\%.

\section{Results}\label{results2}

\begin{figure*}
  \centering
  \resizebox{\hsize}{!}{\includegraphics[bb=35 395 540 760,clip=]{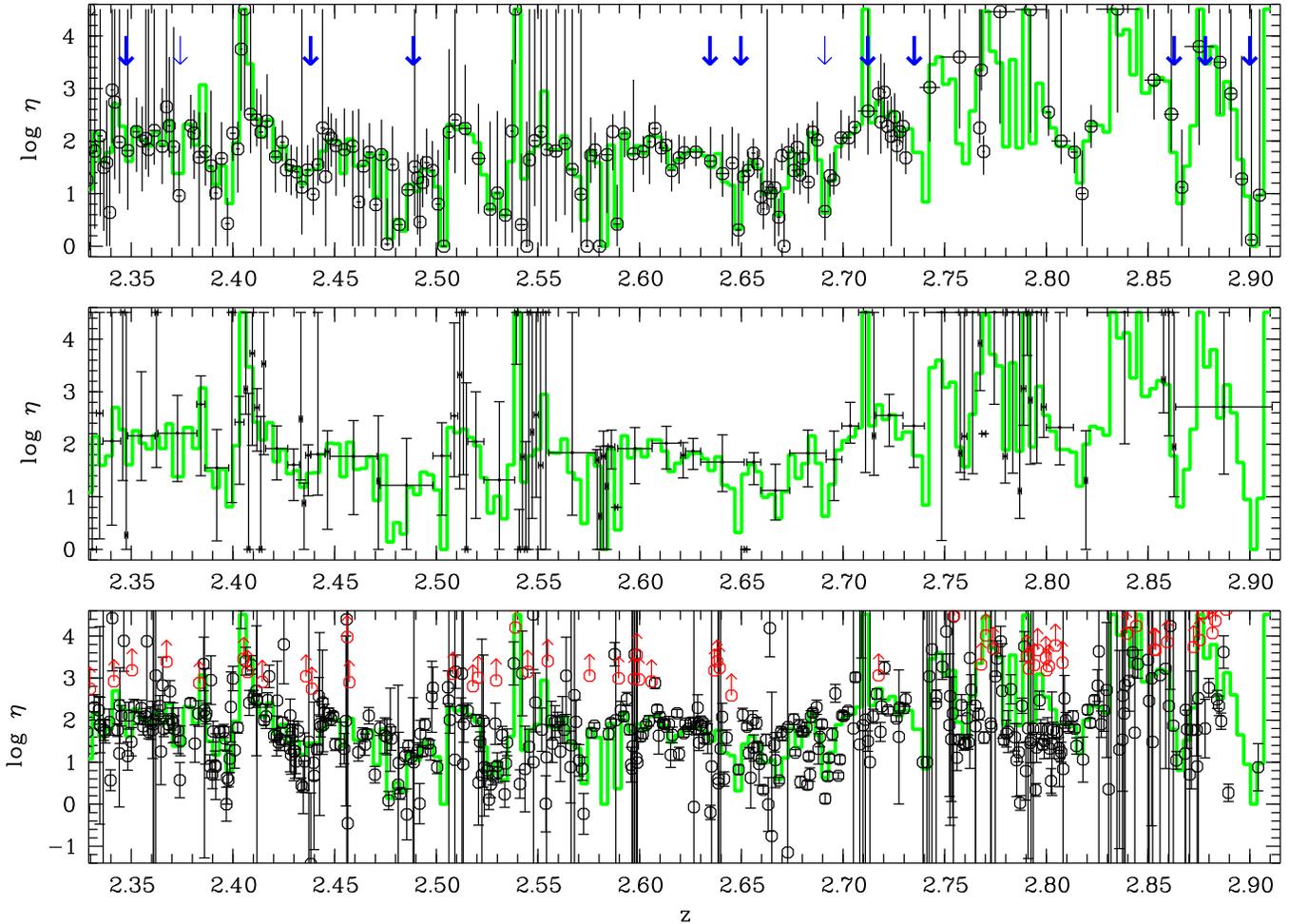}}
  \caption{Distribution of $\eta$ with redshift towards HE~2347-4342 defining the fit interval by eye (upper panel) or using a $\chi^2$ procedure (middle panel) as well as the result from profile fitting (lower panel). For a better orientation, the histogram-like line in each panel indicates $\eta$ estimated with a constant $1.0\,\mathrm{\AA}$ bin size.
The redshifts of metal line absorption systems are marked as arrows in the upper panel.
  }
  \label{all}
\end{figure*}

Applying the four scale-defining methods described above, the data are analyzed.
For comparison, we also perform a Doppler profile analysis using the line fitting program CANDALF.
The resulting distributions of $\eta$ with redshift in case of HE~2347-4342 are presented in Fig.\ \ref{all}.
The panels show the results for binning by eye, binning using the $\chi^2$ method, and from the Doppler profile analysis. 
For a better orientation, the result for constant binning of $1.0\,\mathrm{\AA}$ bin size are shown as overlay in each panel.

The low $S/N$ of the FUSE data leads to large error bars, regardless of how the fit interval is defined.
At longer wavelengths ($\gtrsim 1135\,\mathrm{\AA}$ corresponding to $z \gtrsim 2.73$), instead of a forest structure, the \ion{He}{ii} absorption gets ``patchy''.
This means that regions of high opacity and nearly complete absorption change with opacity gaps and continuum windows \citep{reimersetal1997, smetteetal2002}.
Unlike the Ly$\alpha$ forest, the absorption in the patchy zone is thought to arise from a medium not yet fully reionized \citep{reimersetal1997}.
Therefore, different results may be expected and we distinguish between the Ly$\alpha$ forest visible in \ion{He}{ii} and the patchy absorption towards HE~2347-4342. 
The slightly less redshifted QSO HS~1700+6416 probes only material that is already completely reionized. 
This means, solely Ly$\alpha$ forest absorption is detected.
In the following, we first present the results from the Ly$\alpha$ forest 
in the spectrum of HE~2347-4342 and thereafter HS~1700+6416.

\subsection{The Ly$\alpha$ forest towards HE~2347-4342}

\begin{figure}
  \centering
  \resizebox{\hsize}{!}{\includegraphics[bb=45 325 355 730,clip=]{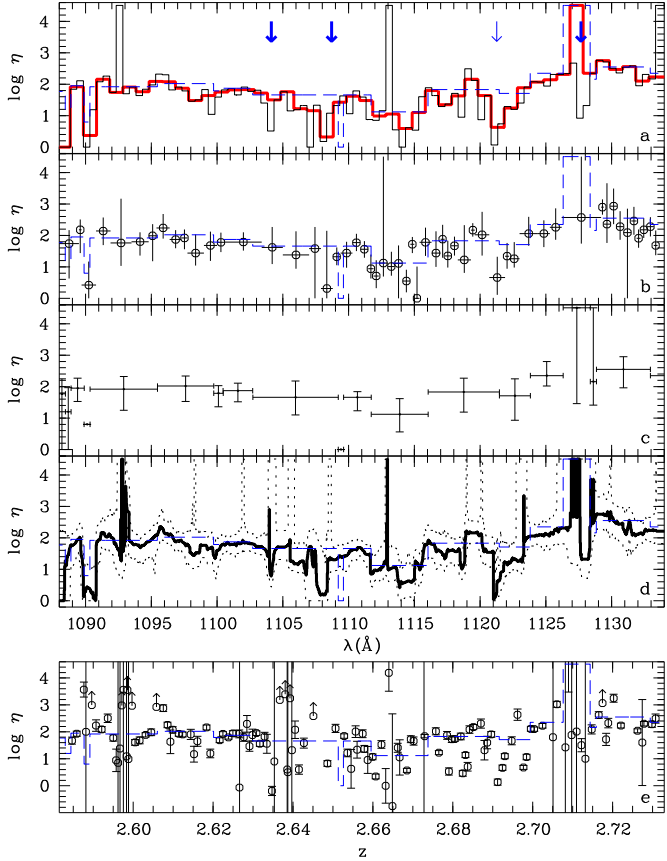}}
  \caption{Redshift distribution of $\eta$ for the best part of the FUSE data ($1088 - 1134\,\mathrm{\AA}$) of HE~2347-4342 using different fit interval definitions.
({\bf a}): constant binning with bin sizes of $0.5\,\mathrm{\AA}$ (thin line) and $1.0\,\mathrm{\AA}$ (thick line; corresponding to the reference distribution in each panel of Fig.\ \ref{all}). The uncertainties are omitted for a clearer presentation.
({\bf b}): fit interval definition by eye.
({\bf c}): bin size estimated by a $\chi^2$ procedure.
({\bf d}): estimation of $\eta$ per pixel averaged over $1.0\,\mathrm{\AA}$, i.e.\ 40 pixels. The dotted line represents the $1\,\sigma$ uncertainties.
({\bf e}): result from profile fitting.
The dashed lines in each panel indicate the $\eta$ distribution yielded with the $\chi^2$ procedure (panel {\bf c}) for better comparability.
All presented errors specify $1\,\sigma$ uncertainties.
The positions of metal absorption systems are indicated as arrows in panel {\bf a}.
  }
  \label{1088-1134}
\end{figure}

As Ly$\alpha$ forest towards HE~2347-4342 we consider the wavelength range $< 1134\,\mathrm{\AA}$ or $z < 2.735$, respectively.
As can be seen from Fig.\ \ref{all}, the $\eta$ values in the redshift range $2.58 < z < 2.73$ are characterized by comparably small error bars.
A close up to this range, where the data quality of the FUSE spectrum is best ($1088 - 1134\,\mathrm{\AA}$), is shown in Fig.\ \ref{1088-1134}.
In this figure the results from all different methods concerning the choice of the scale are presented.
The fact that we find similar distributions of $\eta$ with redshift for all scale definitions (panels a -- d), gives confidence that the choice of the length scale has a minor influence on the results and can, in principle, be derived from the data.
Even the column density ratios of the independent profile-fitting analysis are distributed similarly (Fig.\ \ref{1088-1134}e).
This is noticeable in particular in the redshift regions $2.60 - 2.62$, $2.645 - 2.66$, and $2.68 - 2.70$.

Extremely large error bars of the column density ratios obtained by profile fitting appear to be correlated to extreme $\eta$ values or large uncertainties in the corresponding results of the spectrum fitting procedure. 
Obviously, there are problems with the data at $\sim 1093\,\mathrm{\AA}$, where several noise peaks in the FUSE data mimic an absorption structure different from that of \ion{H}{i} (see also Fig.\ \ref{example}).
Strong saturation, i.e.\ nearly complete absorption over roughly $4\,\mathrm{\AA}$, lead to large error bars at $\sim 1127\,\mathrm{\AA}$.
A similar configuration produces the large uncertainties of the results from line profile fitting at $\sim 1105\,\mathrm{\AA}$ although the spectrum fitting results remain nearly unaffected.

The median value for this best region is $\log\eta \approx 1.75$ in case of the spectrum fitting method, similar for all adopted scales, and $1.80$ in case of the profile fits considering also lower limits.
This is slightly less than the mean value of $\eta = 80$ found by \citet{krissetal2001}.
The $1\,\sigma$ scatter related to the average is roughly $0.7\,\mathrm{dex}$ and slightly larger ($\sim0.8\,\mathrm{dex}$) for the whole Ly$\alpha$ forest observed.
Furthermore, almost all derived $\eta$ values are below $\log\eta = 3.0$.
Since very high $\eta$ values are produced by soft radiation such as stellar sources, this might indicate that even though galaxies contribute to the ionizing radiation, they are not dominating the intergalactic UV background at the observed epoch.

Values of $\eta$ significantly below the mean are visible at \ion{He}{ii} wavelengths around $1090\,\mathrm{\AA}$, $1104\,\mathrm{\AA}$, $1108\,\mathrm{\AA}$, and $1121\,\mathrm{\AA}$ in Fig.\ \ref{1088-1134}.
In addition a local minimum is located at $1128\,\mathrm{\AA}$.
Four of these minima coincide with metal line absorption systems at $z = 2.6346$, $2.6497$, $2.6910$, and $2.7121$ (marked with arrows in Fig.\ \ref{1088-1134}a and summarized in Table \ref{mls2}) associated with saturated \ion{H}{i} absorption features.
The $\eta$ minimum at $1090\,\mathrm{\AA}$ is due to a \ion{H}{i} line ($\log N \approx 13.0$), which seems to mismatch the absorption features in \ion{He}{ii} (see Fig.\ \ref{example}), probably an effect of the poor data quality at this spectral region.
The metal line systems show weak \ion{C}{iv} absorption ($\log N \lesssim12.8$), prominent \ion{O}{vi} features, and no low ionization species.
Since the \ion{O}{vi} features are located in the Ly$\alpha$ forest, their identification might be questionable due to blending with forest lines in at least one doublet component.
Since the systemic redshift can be fixed by the unblended \ion{C}{iv}, the existence of those systems is confirmed.
However, the system at $z=2.6910$ shows no \ion{C}{iv} absorption and the $\lambda 1038$ component of \ion{O}{vi} is located in the Ly$\beta$ absorption trough of the LLS at $z \approx 2.735$. 
Thus the only evidence for metal line absorption is provided by the $\lambda 1032$ component of \ion{O}{vi}.
Since the detected feature is narrow, $b = (11.06 \pm 0.81)\,\mathrm{km\,s}^{-1}$, it certainly originates from a metal ion although an alternative identification cannot be ruled out completely.

\begin{table}
  \caption[]{Metal line systems with $z > 2.3$ in the spectrum of HE~2347-4342.
  }
  \label{mls2}
  $$ 
  \begin{array}{c l }
    \hline
    \hline
    \noalign{\smallskip}
    z & \mathrm{observed~ions}  \\
    \noalign{\smallskip}
    \hline
    \noalign{\smallskip}
    2.3132^{\mathrm{a}} & \ion{C}{iii},~\ion{C}{iv},~\ion{Si}{iii},~\ion{Si}{iv}\\
    2.3475 & \ion{C}{iii}^{\mathrm{c}},~\ion{C}{iv},~\ion{O}{vi}^{\mathrm{b}}\\
    2.3741 & \ion{O}{vi}^{\mathrm{b}}\\
    2.4382 & \ion{C}{iii}^{\mathrm{c}},~\ion{C}{iv},~\ion{O}{vi}^{\mathrm{c}}\\
    2.4887 & \ion{C}{iv},~\ion{O}{vi}^{\mathrm{c}}\\
    2.6346 & \ion{C}{iv},~\ion{O}{vi}\\
    2.6497 & \ion{C}{iii}^{\mathrm{c}},~\ion{C}{iv},~\ion{O}{vi}\\
    2.6910 & \ion{O}{vi}^{\mathrm{a}}\\
    2.7121 & \ion{C}{iv},~\ion{O}{vi}^{\mathrm{a}}\\
    2.735^{\mathrm{d}} & \ion{C}{ii},~\ion{C}{iii},~\ion{O}{vi},~\ion{Si}{iii},~\ion{Si}{iv}\\
    2.8628 & \ion{C}{iii},~\ion{C}{iv},~\ion{O}{vi}\\
    2.8781 & \ion{C}{iv},~\ion{N}{v} \\
    2.895^{\mathrm{e}} & \ion{C}{iii},~\ion{C}{iv},~\ion{N}{v},~\ion{O}{vi} \\
    \noalign{\smallskip}
    \hline
  \end{array}
  $$ 
\begin{list}{}{}
  \item[$^{\mathrm{a}}$] apparently related to a QSO close to the line of sight at $z = 2.31$ (Jakobsen, private communication)
  \item[$^{\mathrm{b}}$] only one doublet component detected
  \item[$^{\mathrm{c}}$] uncertain detection due to blends
  \item[$^{\mathrm{d}}$] Lyman limit system
  \item[$^{\mathrm{e}}$] complex associated system
\end{list}
\end{table}

Further apparent minima of $\eta$ are found at lower redshift.
The position of metal absorption systems are indicated as arrows in the upper panel of Fig.\ \ref{all}.
The four metal line systems with $2.33 < z < 2.50$ showing only highly ionized species coincide again with minima of $\eta$. 
The system at $z = 2.4887$ appears to be located in between of two low $\eta$ regions according to the upper panel of Fig.\ \ref{all}.
However, referring to the lower panels, only the minimum at $z \sim 2.48$ may be real. 

Additional low $\eta$ regions are visible at $z \sim 2.39$ and $2.53$.
The latter might be affected by the extremely high noise in the corresponding \ion{He}{ii} wavelength range, while the first one is an apparent void in \ion{H}{i}.
Within $\Delta z \approx 0.063$ ($2.368 \lesssim z \lesssim 2.430$) we only detect lines with $\log N_{\ion{H}{i}} \lesssim 13.5$.
At the central position of the void ($z \approx 2.398$), two stronger lines ($\log N_{\ion{H}{i}} \approx 13.5$) arise.
We suspect that this void could be due to a quasar close to the line of sight, whose hard radiation field ionizes \ion{H}{i} and \ion{He}{ii} resulting in the observed low $\eta$. 
Fitting the range $2.388 \lesssim z \lesssim 2.401$ yields $\log\eta = 1.72^{+1.19}_{-1.11}$.
The appearance of the fit suggests fluctuations on scales smaller than $\Delta z \approx 0.013$ also reflected in the large $1\,\sigma$ errors.

\begin{figure}
  \centering
  \resizebox{\hsize}{!}{\includegraphics[bb=35 570 355 730,clip=]{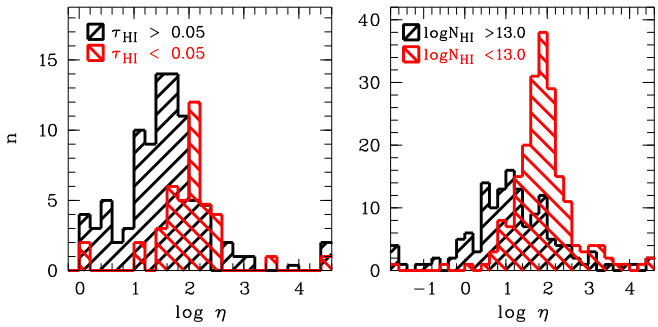}}
  \caption{Distribution of $\eta$ in voids and filaments derived from binning with constant bin size ($1\,\mathrm{\AA}$; left panel) and line profile analysis (right panel) in the Ly$\alpha$ forest region ($z \lesssim 2.73$) towards HE~2347-4342.
Following \citet{shulletal2004}, voids are defined as bins with a mean opacity in hydrogen of $\tau_{\ion{H}{i}} < 0.05$. Concerning the results from the line profile-fitting procedure, we consider lines with $\log N_{\ion{H}{i}} > 13.0$ to arise from filaments.
}
  \label{void_correlation}
\end{figure}

\begin{figure*}
  \centering
  \resizebox{\hsize}{!}{\includegraphics[bb=35 395 545 760,clip=]{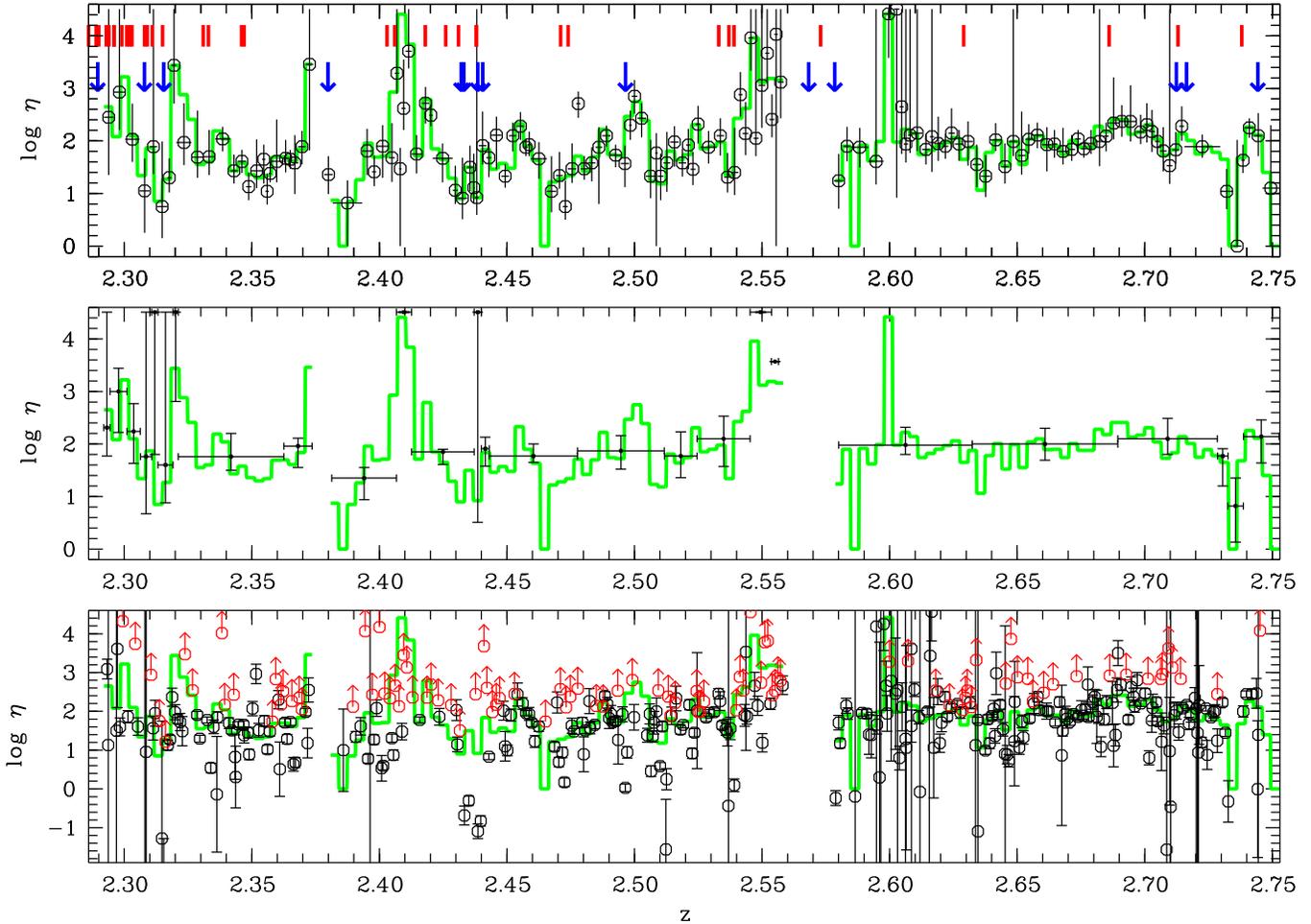}}
  \caption{Redshift distribution of $\eta$ towards HS~1700+6416 according to the spectrum fitting method defining the fit interval by eye (upper panel) or using a $\chi^2$ procedure (middle panel) as well as the result from profile fitting (lower panel). For a better orientation, the histogram-like line in each panel indicates $\eta$ estimated with a constant $1.0\,\mathrm{\AA}$ bin size.
The redshifts of metal line absorption systems are marked as arrows in the upper panel, the position of galaxies close to the line of sight from the compilation of \citet{shapleyetal2005} are indicated by thickmarks as well.
  }
  \label{hs1700}
\end{figure*}

\citet{shulletal2004} found a correlation between the strength of the \ion{H}{i} absorption and $\eta$, in the sense that higher $\eta$ values are measured in \ion{H}{i} voids.
This finding was confirmed by \citet{reimersetal_fuse} towards HS~1700+6416.
We re-investigate this point estimating the mean \ion{H}{i} opacity of the considered fit intervals.
The distribution of $\eta$ for constant bins of $1.0\,\mathrm{\AA}$ size of high ($\tau_{\ion{H}{i}} > 0.05$) and low ($\tau_{\ion{H}{i}} < 0.05 $) opacity is shown in the left panel of Fig.\ \ref{void_correlation}.
Again, only the Ly$\alpha$ forest range is considered.
The mean values are $\log\eta = 1.53 \pm 0.79$ and $1.98 \pm 0.71$ in the high and low opacity sample, respectively. 
For comparison, the corresponding distribution obtained from the profile-fitting procedure is presented in the right panel of Fig.\ \ref{void_correlation}.
The optical depth of $\tau = 0.05$ at the central point of the Doppler profile for a \ion{H}{i} Ly$\alpha$ line of the width $b = 27.0\,\mathrm{km\,s}^{-1}$ is reached at $\log N = 12.25$.
In Ly$\alpha$ forest statistics, a void is defined as a region without any absorption features stronger than $\log N_\ion{H}{i} \sim 13.5$ over a certain comoving size \citep[e.g.][]{kimetal2001}.
As a compromise and in order to maintain a sufficient number of high density absorbers, we consider lines with $\log N < 13.0$ as low density absorbers.
Furthermore, lines with column densities $\lesssim 10^{13}\,\mathrm{cm}^{-2}$ seem to remain unaffected by effects of thermal line broadening (see Sect.\ \ref{discussion3} for more details). 
For the low density sample we find the mean value $\log\eta = 1.90 \pm 0.74$.
Absorbers with $\log N > 13.0$ lead to $\log\eta = 1.21 \pm 1.50$.
Considering different samples of strong \ion{H}{i} absorbers, we find that the mean value of $\log\eta$ decreases with increasing line strength, either defined by optical depth or \ion{H}{i} column density.
This corresponds to the suggestion of \citet{shulletal2004} that low density absorbers correlate with high $\eta$ values. 
It is also consistent with the apparent correlation of metal line systems with strong \ion{H}{i} absorption and low $\eta$ values.
We suspect that this correlation is, at least partly, due to thermal line broadening, which lead to an underestimation of the \ion{He}{ii} column density for absorption features with $\log N_{\ion{H}{i}} \gtrsim 13.0$ if turbulent line widths are assumed.
We will return to this point for further discussion in Sect.\ \ref{discussion3}.

\subsection{HS~1700+6416}

For the analysis of the \ion{He}{ii} data of the quasar HS~1700+6416, we follow the same method as described in Sect.\ \ref{method}.
Additionally, the metal lines predicted to arise in the FUSE spectral range \citep{fechneretal2006a} are considered during the optimization procedure modifying the $\chi^2$-estimation (Eq.\ \ref{chi2}):
\begin{equation}
  \chi^2 = \frac{1}{n}\sum_{i=1}^{n}\frac{\left(f_{\ion{He}{ii},\,i} - \exp\left(-\frac{\eta}{4}\cdot\tau_{\ion{H}{i},\,i}-\tau_{\mathrm{met},\,i}\right)\right)^2}{\sigma_{\ion{He}{ii},\,i}^2 + \sigma_{\ion{H}{i},\,i}^2}\,\mbox{.}
\end{equation}
This strategy is used, since due to the noise the results are more stable if the UV metal lines are considered as part of the fit instead of removing their optical depth from the FUSE data.

Fig.\ \ref{hs1700} presents the redshift distribution of $\eta$ towards HS~1700+6416 derived by spectrum fitting defining the fit interval by eye (upper panel), using a $\chi^2$ procedure (middle panel), or adopting profile fits (lower panel).
Metal lines and interstellar absorption have been considered when fitting Doppler profiles  \citep[for details see][]{fechneretal2006b}.
In the redshift range $2.60 < z < 2.75$ there is apparently only little scatter in $\eta$.
Indeed, the whole range ($2.5815 \le z \le 2.7527$, corresponding to $\Delta z = 0.1712$) can be fitted with $\log\eta = 1.98^{+0.36}_{-0.31}$, which is in good agreement with the mean and median values resulting from the applied methods.
The low value of $\eta$ at $z \approx 2.735$ is due to continuum in \ion{H}{i} and \ion{He}{ii}.
In this case, the applicability of the spectrum fitting method is limited and it tends to produce low $\eta$ values.

Stronger fluctuations are visible in the lower redshift range although the search for a redshift scale using $\chi^2$-estimation leads to smooth variation around $\log\eta \sim 1.8$ at $z = 2.479 \pm 0.066$.
Fitting the whole range yields $\log\eta = 1.78^{+0.43}_{-0.49}$.
High $\eta$ values are apparently found at $z \approx 2.41$.
This peak is most likely due to strong interstellar \ion{C}{ii} $\lambda 1036$ absorption, which is not considered in the spectrum fit procedure.
Furthermore, strong \ion{C}{iv} absorption from the Lyman limit system at $z = 2.3155$ is expected in the UV, which is probably underpredicted \citep[see][]{fechneretal2006a}.
Therefore, we overestimate the amount of absorption due to \ion{He}{ii} leading to unrealistically high values of $\eta$.
Considering the interstellar \ion{C}{ii} absorption should reduce the derived values.
Compared to the result from the Doppler profile fits, which includes the galactic \ion{C}{ii} line, no apparent high $\eta$ peak is visible in agreement with the above arguments.

\begin{figure}
  \centering
  \resizebox{\hsize}{!}{\includegraphics[bb=35 600 280 755,clip=]{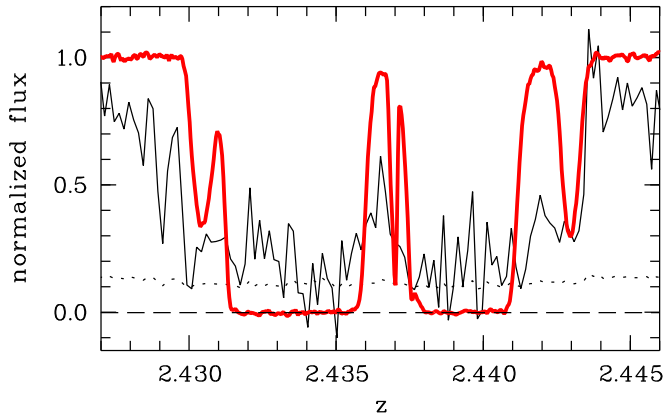}}
  \caption{Absorption features of \ion{H}{i} Ly$\alpha$ (thick line) and \ion{He}{ii} Ly$\alpha$ (thin line) at $z \sim 2.435$. 
The profile of the \ion{He}{ii} features indicates saturation, even though the flux does not reach zero. 
The dotted line represents the error of the \ion{He}{ii} flux. 
The narrow line in between both the saturated \ion{H}{i} troughs is \ion{Si}{ii} of the LLS at $z = 2.3155$.
  }
  \label{zeroflux}
\end{figure}

Close to the metal line system at $z = 2.4965$, another region of increased $\eta$ can be seen, where the redshift distribution yielded by profile fitting resembles that found with the spectrum fit.
The most prominent \ion{H}{i} Ly$\alpha$ feature is the central component in a region ($2.485 \lesssim z \lesssim 2.504$) of weak \ion{H}{i} absorption, i.e.\ a void.
In the corresponding \ion{He}{ii} spectral range strong absorption is detected.
However, several data points of very low $\eta$ in the range $z \approx 2.50 - 2.51$ derived from the profile fits reflect a major problem in the interpretation of the results concerning the zero flux level of the data.
The \ion{H}{i} features at this position appear stronger than the corresponding \ion{He}{ii} features.
The same effect is also visible at $z \sim 2.31$, $2.435$, and $2.46$, where the \ion{H}{i} absorption at the two first redshifts is particularly strong, since they are the Ly$\alpha$ features of Lyman limit systems.
As can be seen from Fig.\ \ref{zeroflux}, the strong \ion{He}{ii} absorption features at $z \sim 2.435$ appear to be saturated, even though the line cores do not reach zero.
Therefore, conclusions drawn from the low redshift portion of the FUSE data are only preliminary.
However, in the longer wavelength part of the data ($>1087\,\mathrm{\AA}$, above the FUSE detector gap), no indications for an incorrect zero flux level are found.
Thus, results from the range $z > 2.58$ can be assumed to be more reliable.

In the spectrum of HS~1700+6416 25 metal line systems can be identified \citep[e.g.][]{vogelreimers1995, koehleretal1996, fechneretal2006a}.
13 of them are at redshifts $z > 2.29$, i.e.\ their \ion{He}{ii} absorption is located in the observed FUSE spectral range.
However, three are affected by the \ion{H}{i} Ly$\beta$ airglow emission or the detector gap.
From the remaining systems, only three arise in the reliable longer wavelength region.
All these are associated to the QSO showing absorption features of \ion{O}{vi} and \ion{N}{v} \citep{simcoeetal2002, fechneretal2006a}.
The system at $z = 2.7443$ is blueshifted with respect to the quasars' emission redshift and shows multicomponent \ion{O}{vi} absorption \citep{simcoeetal2002}.
Since these systems are certainly affected by the quasars' radiation, they are highly suitable to investigate the effect of hard ionizing radiation on $\eta$.
Nevertheless, no indications for a proximity effect are noticeable in the distribution of $\eta$ in Fig.\ \ref{hs1700}.
A slight dip might be present at $z \approx 2.71$ close to the metal line systems also visible in the profile-fitting result.
Considering the blueshifted system, we do not find any hints for a depression of $\eta$.

The absence of a proximity effect in $\eta$ seems to be in conflict with the presence of associated metal line absorption obviously exposed to hard quasar radiation.
However, the observed redshifts of the metal line systems may not be solely cosmological.
This means, the absorbers may be located very close to the QSO ($\sim 10\,\mathrm{kpc}$), ejected from the central region with high velocities.
The system redshifted with respect to the QSO emission $z_{\mathrm{em}} \approx 2.72$ is supposed to trace infalling material, i.e.\ the distance to the central engine is unknown.
Due to the short duty cycle of quasars \citep[$\sim 10^{6} - 10^{7}\,\mathrm{yr}$, e.g.][]{jakobsenetal2003, schirberetal2004}, the assumption of photoionization equilibrium might not be fulfilled for these systems.
In addition, HS~1700+6416 is variable in the intrinsic EUV \citep{reimersetal2005b}.
However, short time variations are not supposed to cause deviations from ionization equilibrium. 

However, there seems to be an apparent correlation of metal line systems and regions of low $\eta$ at redshifts $< 2.55$ (see Fig.\ \ref{hs1700}), which have to be interpreted carefully due to the discussed uncertainties in the zero flux level.

\citet{fechneretal2006a} found that the majority of the metal line systems with $z \gtrsim 2$ towards HS~1700+6416 can be modelled convincingly using a modified version of the \citet{haardtmadau2001} UV background.
The modification of the shape of the ionizing spectrum is a shift of the break usually located at $4\,\mathrm{Ryd}$ to lower energies ($3\,\mathrm{Ryd}$, HM3).
The column density ratio $\eta$ expected for this radiation field is $150 - 190$, depending on the exact position of the break.
If this shape of the ionizing radiation is representative for the IGM in general, the expected column density ratio would be $\log\eta \sim 2.2$, consistent with the value found by fitting the \ion{H}{i} spectrum to the upper FUSE spectral range.
The metal line systems, however, probe the high density IGM.
Therefore, according to the analysis of metal line systems, at least the high density absorbers are expected to show high $\eta$ values, which contradicts the results obtained by measuring \ion{H}{i} and \ion{He}{ii} directly.
We suspect that this conflict can be solved if thermal line broadening is taken into account.
The distribution of the Doppler parameters derived from line profile fitting has its maximum at $b_{\ion{H}{i}} \approx 23\,\mathrm{km\,s}^{-1}$.
If this line width is interpreted to be completely thermal, the corresponding temperature is roughly $T \approx 3\cdot 10^{4}\,\mathrm{K}$, which is close to the value $2\cdot 10^{4}\,\mathrm{K}$ estimated by \citet{ricottietal2000}.
\citet{laietal2006} even predict temperature fluctuations due to inhomogeneous \ion{He}{ii} reionization with temperatures as high as $3\cdot 10^{4}\,\mathrm{K}$.
Thus, contributions from thermal broadening to the line width are expected to be important at least for part of the absorbers.

\section{Discussion}\label{discussion3}

\begin{figure}
  \centering
  \resizebox{\hsize}{!}{\includegraphics[bb=40 425 330 740,clip=]{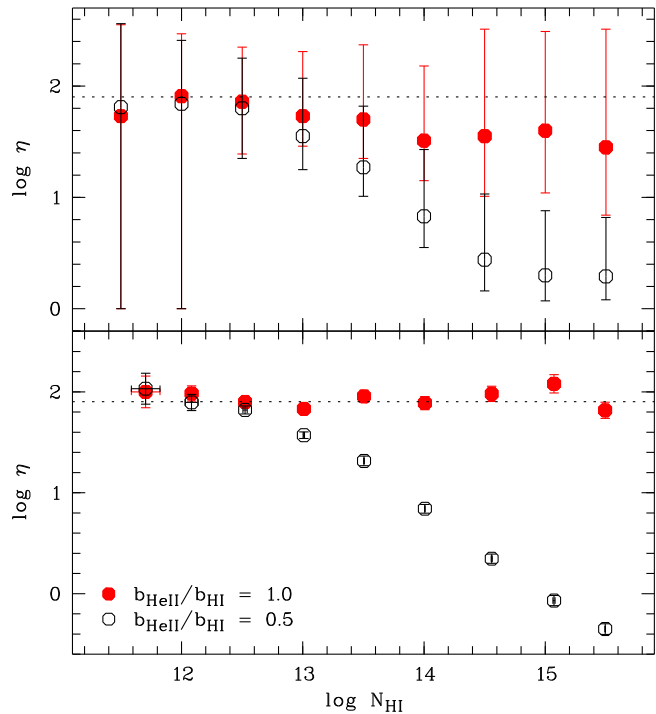}}
  \caption{Dependence of $\eta$ on the strength of the \ion{H}{i} absorption as derived from artificial absorption features in case of the spectrum fitting method (upper panel) and the profile-fitting procedure (lower panel). 
For several column densities of \ion{H}{i} and $b_{\ion{H}{i}} = 27.0\,\mathrm{km\,s}^{-1}$, line profiles of \ion{H}{i} and \ion{He}{ii} are calculated with $\eta = 80$ and $b_{\ion{He}{ii}} = b_{\ion{H}{i}}$ (turbulent broadening, filled circles) or $b_{\ion{He}{ii}} = 0.5\cdot b_{\ion{H}{i}}$ (thermal broadening, open circles), respectively. 
The artificial data was analyzed assuming pure turbulent broadening.
The dashed line indicates the presumed value of $\eta$. 
  }
  \label{turbtherm}
\end{figure}

The analysis of the \ion{He}{ii} data of HE~2347-4342 revealed an apparent correlation between regions of low $\eta$ and the presence of metal line absorption systems.
This finding is confirmed towards HS~1700+6416, albeit systematical uncertainties possibly affect the results of this line of sight.
In order to investigate biases caused by the assumption made for the investigation, we analyze artifical data.
Since turbulent line broadening is implicitly assumed in the spectrum fitting procedure, we consider two sets of profiles dominated by thermal and turbulent broadening, respectively.

The artificial lines are computed as Doppler profiles with $z=2.6194$, $b_{\ion{H}{i}} = 27.0\,\mathrm{km\,s}^{-1}$ and different values of the \ion{H}{i} column density with $11.5 \le \log N_{\ion{H}{i}} \le 15.5$.
The \ion{He}{ii} column density is $\log N_{\ion{He}{ii}} = \log\eta+\log N_{\ion{H}{i}}$ with $\eta = 80$.
Two \ion{He}{ii} features are computed for each column density representing both the extremes of line broadening, pure turbulence $b_{\ion{He}{ii}} = b_{\ion{H}{i}}$ and thermal broadening $b_{\ion{He}{ii}} = \frac{1}{2}\,b_{\ion{H}{i}}$.
The generated line profiles are convolved to match the resolution of the observed optical ($R \approx 45\,000$) or UV spectra ($R \approx 20\,000$), respectively.
Noise is added to have $S/N \sim 100$ in case of \ion{H}{i} and $\sim 5$ in case of \ion{He}{ii}. 

The simulated lines are analyzed twice, fitting the spectrum and Doppler profiles.
Fig.\ \ref{turbtherm} shows the inferred $\eta$ as a function of the underlying \ion{H}{i} column density.
Both fitting procedures severely underestimate $\eta$ for strong \ion{H}{i} lines if the assumption of turbulent broadening is incorrect.
Here we consider the extreme of pure thermal broadening, where the Doppler parameter of \ion{He}{ii} is half the value of \ion{H}{i}.
In reality, there will be a mixture of thermal and turbulent broadening.
\citet{zhengetal2004} even find that turbulent broadening is dominant.
This finding has been confirmed by \citet{liuetal2006} who compared the observations to cosmological hydrodynamical simulations.
Also \citet{boltonetal2006} argue that thermal broadening is supposed to be negligible, since Hubble broadening is expected to be the dominant contribution to the width of Ly$\alpha$ forest lines.
However, the stronger the contribution of thermal broadening, the more the inferred $\eta$ will underestimate the actual value.

If turbulent broadening is indeed dominating, line profile fitting recovers well the underlying $\eta$.
In this case, the artificial lines are computed as Doppler profiles.
Therefore, profile fits are supposed to recover the correct values, in particular since only a single line is considered.
A more detailed exploration would require hydrodynamical simulations.
Such an investigation is presented by \citet{boltonetal2006} exploring the applicability of line profile fitting.
They find that $\eta$ is inferred accurately using this technique (but see discussion below).
\citet{liuetal2006} additionally investigate the effect of thermal broadening and peculiar velocity fields using hydrodynamical simulations.
They conclude that the absorption features are predominately turbulent broadened but fluctuations of the temperature and the velocity fields contribute to the scatter in $\eta$.

According to our synthetic line profiles, the spectrum fitting method exhibits deviations from the underlying $\eta$ value even in the case of turbulent broadening.
The reason is that we use the relation $\tau_{\ion{He}{ii}} = \eta/4\cdot\tau_{\ion{H}{i}}$ to scale the \ion{H}{i} spectrum.
 If the \ion{H}{i} absorption features are saturated, the accuracy of the optical depth measurement is limited.
Though the correct $\eta$ is recovered within the $1\,\sigma$ error bars, the root-mean-square deviation is  roughly $0.16$ for column densities $\log N_{\ion{H}{i}} \ge 13.5$.
Considering a sample of synthetic line profiles with $\log\eta = 1.0$, the spectrum fit method recovers the underlying column density ratio more accurately in the case of stronger \ion{H}{i} features. 

With this information about potential methodical problems, there are three possible interpretations of the apparent correlation between low $\eta$ and strong \ion{H}{i} absorption: (1) a methodical bias, (2) an effect of thermal line broadening, or (3) physical reality.
In the following, we will discuss these interpretations and their implications.

\begin{figure*}
  \centering
  \resizebox{\hsize}{!}{\includegraphics[bb=45 540 445 725,clip=]{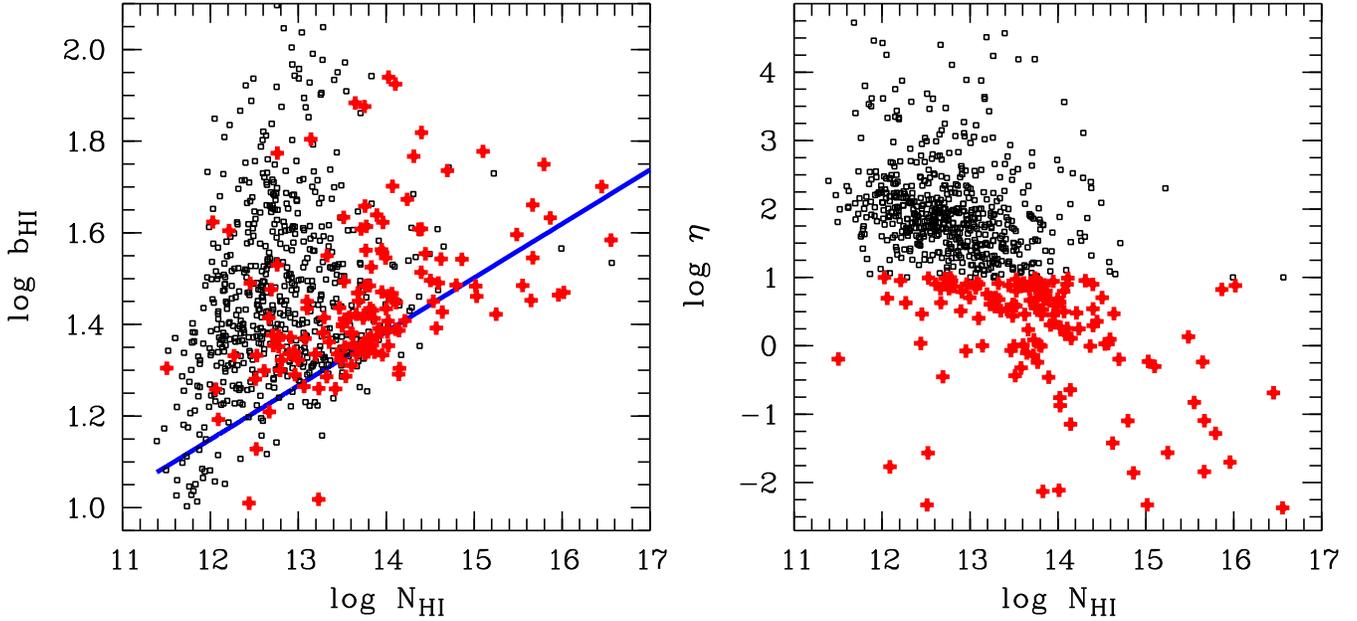}}
  \caption{Measured \ion{H}{i} column density versus logarithmic Doppler parameter (left panel) for the combined line sample of HE~2347-4342 and HS~1700+6416.
The solid line represents the iteratively estimated cut-off $b$-parameter $\log b_{\mathrm{c}} = 0.118\cdot\log N_{\ion{H}{i}} -0.26$ (see text).
Only lines with $12.5 < \log N_{\ion{H}{i}} < 15.0$ and $\sigma_{b}/b < 0.5$ and $\sigma_{\log N_{\ion{H}{i}}}/\log N_{\ion{H}{i}} < 0.5$ are considered.
The right panel shows are the distribution of $\eta$ with the \ion{H}{i} column density.
Lines with $\log\eta < 1.0$ are highlighted.
  }
  \label{logN_b}
\end{figure*}

\subsection{Methodical uncertainties}
Due to methodical effects $\eta$ might be systematically underestimated for strong \ion{H}{i} absorbers, since $\tau_{\ion{He}{ii}} = \eta/4\cdot\tau_{\ion{H}{i}}$ leads to accurate results only for unsaturated lines. 
Since metal lines arise particularly in strong Ly$\alpha$ systems, this would lead to the apparent correlation.
At first glance only the results from spectrum fitting should be affected in this way.
In accordance with Fig.\ \ref{turbtherm} and the investigation of \citet{boltonetal2006} profile fits recover the underlying $\eta$ values.
However, the scatter plots of the column density ratio $\eta$ presented by \citet[][their Fig.\ 4]{boltonetal2006} show a slight tendency to smaller $\eta$ values in the case of higher \ion{H}{i} column densities, even for a uniform UV background.
Due to the power law behaviour of the column density distribution function \citep[e.g.][]{kirkmantytler1997, kimetal2001}, high column density absorbers are less numerous than weaker lines.
Thus, strong absorbers represent only a minor fraction of the investigated line sample with a negligible contribution to the overall statistics.
Therefore, it is necessary to extend the analysis of simulated data with the purpose to obtain a sufficiently large statistical sample in order to investigate high column density absorbers ($\log N_{\ion{H}{i}} \gtrsim 14.0$) in more detail.

\subsection{Thermal broadening}
Assuming that the profile-fitting procedure recovers the column density ratio accurately, the apparent correlation between metal line absorption and low $\eta$ should be no methodical artifact.
The reason is that the correlation is also recognizable in the profile-fitting results (Fig.\ \ref{all} and \ref{hs1700}). 
Then, Fig.\ \ref{turbtherm} suggests that the line widths of \ion{H}{i} and \ion{He}{ii} features with associated metal line absorption might be dominated by thermal broadening.
In general, lines with $N_{\ion{H}{i}} \gtrsim 10^{13}\,\mathrm{cm}^{-2}$ if dominated by thermal broadening, would produce extremely low $\eta$ values due to the method for both procedures.
As a consequence the majority of the lines showing low $\eta$ values from profile fitting should be located close to the cut-off of the $b(N)$ distribution.

Fig.\ \ref{logN_b} shows the measured \ion{H}{i} column density versus the $b$-parameter (both logarithmic) as well as $\log N_{\ion{H}{i}}$ versus the column density ratio $\eta$ for the combined line sample of HE~2347-4342 and HS~1700+6416.
The right panel illustrates the correlation of $\eta$ with the strength of the \ion{H}{i} absorption.
Points with $\log\eta < 1.0$ are highlighted.
Their average column density is $13.80 \pm 0.95$.
Less than 20\,\% of the low $\eta$ lines have \ion{H}{i} column densities $< 10^{13}\,\mathrm{cm}^{-2}$, supporting the suspicion that only absorbers with column densities above this threshold are affected.
In the $N$-$b$ diagram points with $\log\eta < 1.0$ accumulate close to the lower envelope.
This means, thermal broadening is expected to be relevant for these lines. 

We roughly estimate the cut-off $b$-parameter by an iterative procedure \citep{schayeetal1999, schayeetal2000, kimetal2002} including only lines with $\sigma_b/b < 0.5$ and $\sigma_{\log N_{\ion{H}{i}}}/\log N_{\ion{H}{i}} < 0.5$.
According to \citet{huignedin1997}, the tight power law relation between the temperature and density is only valid in the low density IGM, which gives raise to the majority of the Ly$\alpha$ forest features.
Using the relation between hydrogen column density and overdensity $\delta$ given by \citet{schaye2001}, an overdensity of $\delta \approx 10$ roughly corresponds to $\log N_{\ion{H}{i}} \approx 15.0$.
In order to avoid biases due to incompleteness \citep{kimetal2002}, only lines above the completeness limit are considered.
We constrain the sample used for the fit to lines with $12.5 < \log N_{\ion{H}{i}} < 15.0$.
The final sample contains about 66\,\% of the total lines and has a mean redshift of $\langle z \rangle = 2.58$.

For the model $\log b_{\mathrm{c}} = (\Gamma -1)\cdot\log N_{\ion{H}{i}} + \log b_0$ we find $(\Gamma -1) = 0.118 \pm 0.005$ and $\log b_0 =  -0.26 \pm 0.07$ in agreement with the results of \citet{kimetal2002}.
Considering the low $\eta$ lines ($\log\eta < 1.0$), about 70\,\% are within $5\,\sigma$ of the cut-off, where we adopt $\sigma = 0.026$ the standard error of the fit estimate.
Furthermore, the mean column density ratio of all data points of the fit sample within $\log b_{\mathrm{c}}\pm 5\,\sigma$ is $\log\eta = 1.31 \pm 1.08$ (median $1.25$), which is lower than the mean value of the total fit sample $1.56\pm 1.08$ (median $1.60$).
The effect gets more pronounced for data points located closer to the cut-off.
Considering the points within $1\,\sigma$ the median value of $\log\eta$ is $1.07$.
Extrapolating the estimated cut-off to lower and higher column densities, we find the same effect, i.e.\ the median column density ratio decreases, the closer the considered lines are located to the cut-off.
Within $5\,\sigma$ ($2\,\sigma$, $1\,\sigma$) the median value is $\log\eta = 1.45$ ($1.39$, $1.24$).

\begin{figure}
  \centering
  \resizebox{\hsize}{!}{\includegraphics[bb=50 510 330 725,clip=]{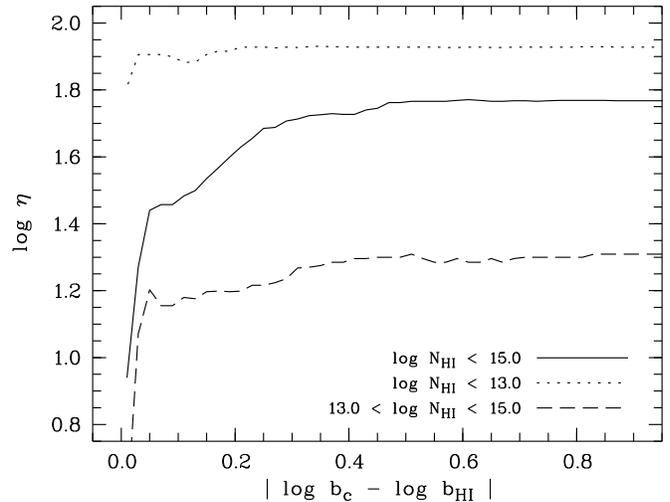}}
  \caption{Cumulative distribution of the median $\log\eta$ with the distance $d = |\,\log b_{\mathrm{c}} - \log b_{\ion{H}{i}}\,|$ from the $b(N)$ cut-off.
The median column density ratio is estimated considering all absorbers within the given distance from the cut-off $\log b_{\mathrm{c}}$.
The solid line represents all absorbers with $\sigma_{b}/b < 0.5$, $\sigma_{\log N_{\ion{H}{}}}/\log N_{\ion{H}{i}} < 0.5$, and $\log N_{\ion{H}{i}} < 15.0$.
Additionally, the distributions of the low density sample containing absorbers with $\log N_{\ion{H}{i}} < 13.0$ (dotted) and the high density sample with lines $13.0 < \log N_{\ion{H}{i}} < 15.0$ (dashed) are indicated.
  }
  \label{cutoff_eta}
\end{figure}

Fig.\ \ref{cutoff_eta} illustrates the dependence of the inferred column density ratio from the distance $d = |\log b_{\mathrm{c}} - \log b_{\ion{H}{i}}|$ to the cut-off in the $b(N)$ distribution, i.e.\ from the contribution of thermal line broadening.
The cumulative distribution of the median $\eta$ values is given for the total ($\log N_{\ion{H}{i}} < 15.0$) sample as well as a low and high density subsample, which comprise the lines with $\log N_{\ion{H}{i}} < 13.0$ and $13.0 < \log N_{\ion{H}{i}} < 15.0$, respectively.
It is clearly seen that the low density sample approaches a larger median $\eta$ value ($\log\eta = 1.93$) than the high density sample ($1.31$).
The low density sample reaches the plateau values close as 98\,\% at $d \approx 0.03$, where only 4\,\% of the data points are used to estimate the median.
The same level, i.e.\ 98\,\% of the plateau value, is reached at $d \approx 0.65$ in the case of the high density sample, where 98\,\% of the absorbers are included.
These numbers illustrate that high density absorption features are statistically more sensitive to the distance the the $b(N)$ cut-off than low density lines.
The cumulative distribution of the total sample reaches a plateau value of 92\,\% of that of the low density sample, i.e.\ $\log\eta = 1.77$, but follows the smoother rise of the high density sample.
98\,\% of the plateau value is reached at $d \approx 0.43$.

We conclude that absorption lines dominated by thermal broadening, i.e.\ located close the the thermal cut-off of the $b(N)$ distribution, reveal systematically low column density ratios $\eta$ when applying standard analysis procedures.
In consequence, the apparent correlation between the strength of the \ion{H}{i} absorption and high $\eta$ values should be reconsidered.
As illustrated in Fig.\ \ref{turbtherm}, the column density ratio of weaker \ion{H}{i} absorbers ($\log N \lesssim 13.0$) will be well recovered even if the assumption of turbulent broadening is incorrect.
While weak \ion{H}{i} lines supposedly represent an unbiased sample (represented by the low column density sample shown in Fig.\ \ref{void_correlation}), strong lines are expected to be biased towards lower $\eta$ values.
Hence, weak lines are expected to show on average higher column density ratios than strong lines (see also Fig.\ \ref{cutoff_eta}).
Thus, the correlation of the column density ratio with the strength of the \ion{H}{i} absorption can be explained by inadequate handling of thermal broadened lines by the applied analysis procedures.


According to this result, only the low density Ly$\alpha$ forest should be used to estimate the column density ratio.
Considering only absorbers with $\log N_{\ion{H}{i}} < 13.0$ the mean $\eta$ value of the combined sample of HE~2347-4342 and HS~1700+6416 is $\log\eta = 1.98 \pm 0.81$ (median $1.93$).
If the patchy zone in the spectrum of HE~2347-4342 is neglected, we yield $\log\eta = 1.89 \pm 0.72$ (median $1.90$).
For the Ly$\alpha$ forest towards HE~2347-4342, we find $1.90 \pm 0.74$ (median 1.88), and towards HS~1700+6416 $1.89 \pm 0.71$ (median $1.93$).
These values are roughly $0.3\,\mathrm{dex}$ higher than those obtained including the high density absorbers.
The spectrum fitting method leads to values roughly $0.2\,\mathrm{dex}$ lower than those derived from the low density samples but are significantly higher than the average values for the total line samples, implying that it is less sensitive to the systematic errors due to thermal line widths.
However, this is only valid on scales $\Delta z \gtrsim 0.1$.
Variations of $\eta$ on smaller scales are possibly affected by methodical uncertainties due to thermal line broadening.

Further support is provided by the photoionization models of the metal line systems in the spectrum of HS~1700+6416 derived by \citet{fechneretal2006a}.
The system at $z = 2.4965$, for example is modeled using a modified \citet{haardtmadau2001} UV background (HM3).
Doppler parameters are computed by comparing the model temperature with the observed $b$-parameter of a well-measured ion \citep[for further details see][]{fechneretal2006a}.
For this system, the velocity ratio $\xi = b_{\ion{He}{ii}}/b_{\ion{H}{i}}$ is $\sim 0.50$, i.e.\ the line width is determined by pure thermal broadening.
The modeled column density ratio is $\log\eta = 2.35$.
Fitting line profiles, we assume pure turbulent broadening and find $\log\eta = 0.03 \pm 0.10$ (clearly seen in the lower panel of Fig.\ \ref{hs1700} at $z \approx 2.5$).
This supports the above argumentation that the column density ratio is severely underestimated if absorbers with thermal line widths are analyzed assuming turbulent line broadening.
Considering all metal line systems modeled by \citet{fechneretal2006a}, the mean velocity ratio is $\xi = 0.59 \pm 0.08$, where the column density of the \ion{H}{i} components is on average $\log N_{\ion{H}{i}} = 15.5 \pm 0.7$.
Thus, according to the photoionization models the line widths of strong metal line systems towards HS~1700+6416 are basically due to thermal broadening and can therefore explain the apparent correlation between low $\eta$ values.

\subsection{Physical implications}
Even if part of the low $\eta$ values can be explained by systematical biases due to thermal broadening, part of them might be real.
In such cases, the coincidence of low $\eta$ values and metal line absorption systems, showing features of highly ionized material, suggest the vicinity of a hard radiation source like an AGN.
Alternatively, the presence of \ion{O}{vi} may also imply the presence of a hot, collisionally ionized gas phase.
The suspected redshifts would be $z \sim 2.40$, $2.48$, $2.63$, $2.65$, $2.69$, as well as $2.82$ and $2.86$ towards HE~2347-4342.
Further support for this assumption is provided by the detected metal lines of highly ionized species (see Table \ref{mls2}).
\citet{boltonetal2006} showed that fluctuations may be due to variations in the number, luminosity and spectral shape of the small number of QSOs.
In our case, the redshift range $z \sim 2.40$ towards HE~2347-4342 is of particular interest. 
It is the center of an apparent void in \ion{H}{i}, where we measure low $\eta$ values. 
We find $\log\eta = 1.07^{+1.13}_{-1.07}$ for $2.39 < z < 2.40$. 
In the optical some weak absorption features with $\log N_{\ion{H}{i}} < 13.5$ are detected.
The redshift ranges $2.376 < z < 2.384$ and $2.401 < z < 2.419$, where only \ion{H}{i} lines with $\log N_{\ion{H}{i}} \lesssim 12.7$ are detected, exhibit high $\eta$ values ($\log\eta = 2.14^{+0.49}_{-0.76}$ and $2.41^{+0.83}_{-0.67}$, respectively).
Besides fluctuations due to statistical properties of the ionizing sources, this configuration may indicate the spatial vicinity of a QSO. 
In this case the $z \sim 2.40$ region towards HE~2347-4342 would be one of the rare cases of a classical transversal proximity effect \citep[e.g.][]{schirberetal2004, croft2004}.
However, since there are no published positions of QSOs close to this line of sight for the required redshift range, its presence cannot be confirmed to date.

Towards HS~1700+6416 the amplitude of the observed fluctuations is apparently smaller, in particular in the redshift range $2.58 < z <2.75$ where the data quality is best.
This may be partly due to the lower noise level of the FUSE data.
HS~1700+6416 was target of deep direct observation campaigns with the aim to identify galaxies close to the line of sight \citep{teplitzetal1998, erbetal2003, shapleyetal2005}.
Among the galaxies identified by \citet{shapleyetal2005} there are three objects showing indications of quasar activity.
They are located at $z \approx 2.293$, $2.333$, and $2.347$, which is at the low edge of our FUSE data. 
We see indeed indications of relatively low $\eta$ regions in this redshift range.
However, due to the poor data quality in this part, this coincidence should be considered very preliminary.

\subsection{Spatial scales of $\eta$ variations}

\begin{figure}
  \centering
  \resizebox{\hsize}{!}{\includegraphics[bb=37 570 432 725,clip=]{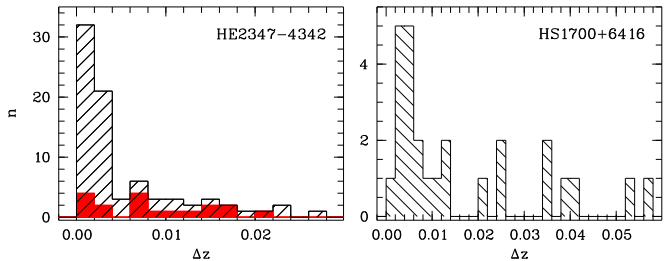}}
  \caption{ Distribution of estimated sizes of the scales of $\eta$ variation towards HE~2347-4342 (left panel; $z < 2.74$) and HS~1700+6416 (right panel).
The left panel also shows the distribution of scale sizes towards HE~2347-4342 considering the best part of the \ion{He}{ii} only (filled histogram; $1088 < \lambda < 1134\,\mathrm{\AA}$ corresponding to $2.58 < z < 2.73$).
  }
  \label{scale_distr}
\end{figure}

\begin{figure}
  \centering
  \resizebox{\hsize}{!}{\includegraphics[bb=35 410 500 700,clip=]{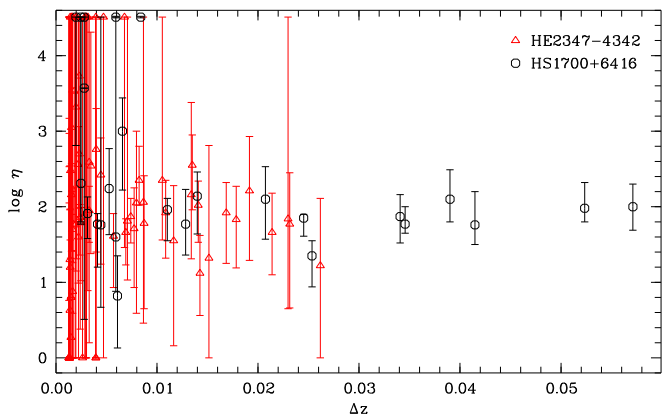}}
  \caption{Scale size versus $\eta$ value for HS~1700+6416 (circles) and the \ion{He}{ii} Ly$\alpha$ forest of HE~2347-4342 (triangles).
  }
  \label{scale_logeta_corr}
\end{figure}

The spectrum fit method leads to estimates of the spatial scales of $\eta$ variations.
Due to the fitting procedure the smallest detectable scales have a size of $\Delta z \approx 0.001$ which corresponds to the value found by \citet{shulletal2004} towards HE~2347-4342.

The resulting distributions of scales as derived by our $\chi^2$-procedure (the middle panels of Figs. \ref{all} and \ref{hs1700}) are shown in Fig.\ \ref{scale_distr}.
Towards HE~2347-4342 only the Ly$\alpha$ forest part $z < 2.73$ of the \ion{He}{ii} spectrum is considered, i.e.\ the patchy not completely reionized part is not used.
In this case scales in the range $0.0013 \lesssim \Delta z \lesssim 0.0473$ are derived.
Obviously, this sample favors small scale variations since 69\,\% of the scales have a size of $\Delta z \le 0.005$, corresponding to $\lesssim 4\,h^{-1}\,\mathrm{Mpc}$ comoving.
Only 20\,\% of the estimated scales are larger than $\Delta z > 0.01$, corresponding to $\gtrsim 8\,h^{-1}\,\mathrm{Mpc}$ comoving. 
Towards HS~1700+6416 larger scales in the range $0.0020 \lesssim \Delta z \lesssim 0.0571$ are found.
For this line of sight 31\,\% of the scales are smaller than $\Delta z < 0.005$.

The discrepancy of the inferred scale sizes between the sight lines may be due cosmic variance.
However, an alternative explanation could be the lower signal-to-noise ratio of the \ion{He}{ii} data of HE~2347-4342 compared to HS~1700+6416.
Limiting the sample to the range $2.58 < z < 2.73$ towards HE~2347-4342 where the quality of the FUSE spectrum is best and the $S/N$ is comparable to the HS~1700+6416 data, leads to a distribution similar to that towards HS~1700+6416 (indicated as the filled histogram in the left panel of Fig.\ \ref{scale_distr}).
Though the limited sample provides only low statistics, we find that 33\,\% of the scales have a size of $\Delta z < 0.005$.
Large scales of $\Delta z > 0.01$ are derived for 39\,\% of the scales (46\,\% towards HS~1700+6416).
This indicates that part of the small scale variations might be mimicked by the high noise level of the \ion{He}{ii} data.
In order to clarify this point spectra with improved $S/N$ (at least by a factor of 2) would be needed.
However, the results presented in this work indicate that a the large
amplitude of the small scale fluctuations would decrease for data with a
higher signal-to-noise ratio.

Differences between small-scale and large-scale variations are illustrated in  Fig.\ \ref{scale_logeta_corr} which presents the distribution of the $\eta$ values with the scale size.
Obviously, the scatter of the $\eta$ values increases with decreasing scale size.
In particular, the $\eta$ values for scales $\Delta z \lesssim 0.005$ towards HE~2347-4342 are extremely uncertain.
The mean $\eta$ value on scales $\Delta z < 0.01$ for the combined sample of both lines of sight is $\log\eta = 2.30 \pm 1.60$ (median $2.05$), while scales of $\Delta z > 0.01$ lead to $\log\eta = 1.86 \pm 0.33$ (median $1.87$).
This means the scatter of $\eta$ for small scales is more than $1\,\mathrm{dex}$ larger than for larger scales.
Considering both quasar sight lines individually leads to the same result.

Thus, we conclude from the analysis of the spectra of HS~1700+6416 and the best part of the HE~2347-4342 data that about 30\,\% of the $\eta$ variations occur on scales $\lesssim 4\,h^{-1}\,\mathrm{Mpc}$ comoving, while roughly 50\,\% of the variations have scales $\gtrsim 10\,h^{-1}\,\mathrm{Mpc}$ comoving.
Large variations of $\eta$ over several orders of magnitude apparently may occur on small scales of a few Mpc size, whereas there is only little scatter of less than $\pm 0.3\,\mathrm{dex}$ in $\eta$ on large scales $\gtrsim 10\,h^{-1}\,\mathrm{Mpc}$ comoving.

\section{Summary and conclusions}\label{conclusions3}

We have re-analyzed the \ion{He}{ii} data of the QSOs HE~2347-4342 and HS~1700+6416, both observed with FUSE.
The new method we applied is based on the idea to fit the observed high-quality optical spectrum to the noisy far-UV data globally.
This approach should minimize effects due to noise, which limits the applicability of the apparent optical depth method \citep{foxetal2005}, and avoid the subjectivity of a profile-fitting procedure. 
Fitting the metal line cleaned \ion{H}{i} Ly$\alpha$ forest to the  \ion{He}{ii} data leads to $\log\eta = 1.79^{+0.71}_{-0.74}$ for $2.33 < z < 2.73$ towards HE~2347-4342 and $\log\eta = 1.87^{+0.41}_{-0.37}$ for $2.29 < z < 2.75$ towards HS~1700+6416.
These results are in agreement with the median values from the Doppler profile analysis, however, slightly lower than the median for lines with $\log N_{\ion{H}{i}} < 13.0$, which we interpret as due to the neglect of thermal broadening (see below).
The mean value of the Ly$\alpha$ forest lines with $\log N_{\ion{H}{i}} < 13.0$ in both quasars is $\log\eta = 1.89 \pm 0.72$, i.e.\ $\eta \sim 80$.

Strong local deviations from the average suggest that the column density ratio is fluctuating on smaller scales.
In order to determine the redshift scale on which $\eta$ fluctuates, we estimate the width of the spectral range that can be fitted with a minimal $\chi^2$.
We find redshift intervals of the sizes $0.0013 \lesssim \Delta z \lesssim 0.0473$ in the case of HE~2347-4342.
Due to the noisy FUSE data, the result may be biased towards small scales.
For the best part of the spectrum ($1088 - 1134\,\mathrm{\AA}$), the average redshift interval is $\Delta z = 0.0082 \pm 0.0065$ (median $0.0072$, i.e.\ $\sim 5.9\,h^{-1}\mathrm{Mpc}$ comoving).
In case of  HS~1700+6416, the typical length scale is $\Delta z = 0.0140$ (median) corresponding to $\sim 11.4\,h^{-1}\,\mathrm{Mpc}$ comoving, but also scales up to $\Delta z = 0.0571$ are found.
However, strong \ion{H}{i} lines, whose line widths are dominated by thermal line broadening, may lead to an apparent decrease of $\eta$.
Consequently, the derived scales are possibly biased towards small values.
A rough estimate neglecting the small values leads to $\Delta z \sim 0.01 - 0.03$, corresponding to $8 - 24\,h^{-1}\mathrm{Mpc}$ comoving. 
In more detail, we find that roughly 30\,\% of the $\eta$ variations take place on small scales $\lesssim 4\,h^{-1}\,\mathrm{Mpc}$ comoving with an amplitude of about $\pm 1.5\,\mathrm{dex}$.
While on large scales $\gtrsim 10\,h^{-1}\,\mathrm{Mpc}$ comoving there is only little scatter of $\eta$ ($\pm 0.3\,\mathrm{dex}$ or even less).

Towards HE~2347-4342 we detect a \ion{H}{i} void in the range $2.37 \lesssim z \lesssim 2.43$, also showing fluctuations of $\eta$ on scales of $\Delta z \sim 0.01 - 0.02$ ($8.5 - 17.0\,h^{-1}\,\mathrm{Mpc}$ comoving).
The void`s central region at $z \sim 2.40$ exhibits low $\eta$ values suggesting the vicinity of a QSO close to the line of sight.
On the other hand, no proximity effect can be seen towards HS~1700+6416, even though there are metal absorption systems showing material like \ion{N}{v} and \ion{O}{vi} supposed to be exposed to the quasar's radiation field. 

The correlation between \ion{H}{i} voids and small $\eta$ values as claimed by \citet{shulletal2004} and \citet{reimersetal_fuse} has been re-investigated.
We find that the distribution of the column density ratio $\eta$ peaks at higher values for samples consisting of stronger \ion{H}{i} absorption lines or higher \ion{H}{i} optical depth, respectively.
However, simple tests with artificial data reveal that both analysis procedures considered here would severely underestimate $\eta$ in the case of strong \ion{H}{i} absorption ($\log N \gtrsim 13.0$) if the line width would be dominated by thermal broadening.
Investigating the $b(N)$ distribution of the line sample obtained from Doppler profile fitting, we find indeed that the majority (70\,\%) of lines with $\log\eta < 1.0$ is located within $5\,\sigma$ of the temperature cut-off.
Inversely, the average column density ratio decreases if the considered line sample is located closer to the cut-off. 
The median value is $\log\eta = 1.32$ ($1.07$) for lines within $\pm 2\,\sigma$ ($1\,\sigma$) from the cut-off.
Photoionization models of the metal line systems towards HS~1700+6416 \citep{fechneretal2006a} support this finding, since there are indications that the line widths of most of the modeled systems are dominated by thermal broadening.
Thus, at least part of the low $\eta$ values are due to systematic biases caused by analyzing thermal broadened absorbers under the assumption of pure turbulent broadening.
Since the sensitivity of the absorbers to this effect is correlated with the \ion{H}{i} column density, thermal line broadening provides an explanation of the apparent correlation between $\eta$ and the strength of the \ion{H}{i} absorption. 
Consequently, only low density \ion{H}{i} absorbers provide an unbiased sample, since their inferred results are insensitive to the assumed line broadening mechanism.  
As median column density ratios for absorbers with $\log N_{\ion{H}{i}} < 13.0$, we find $\log\eta = 1.93$ towards both quasars HE~2347-4342 and HS~1700+6416.
These values are about 10\,\% higher in comparison to those inferred from the total sample.

Metal line absorption usually arises from high density \ion{H}{i} absorbers.
Consequently, they lead systematically to low $\eta$ values.
This is particularly seen in the spectrum of HE~2347-4342.
Towards this QSO the concerned systems show highly ionized species like \ion{C}{iv} and \ion{O}{vi}.
The presence of high ionization lines suggests that the apparent correlation is not only a systematic effect due to the analysis method but the absorbers may be exposed to the hard radiation of local quasars.
However, in order to verify this supposition, a search for QSOs in the field of HE~2347-4342 is required.

\begin{acknowledgements}
We thank the referee for valuable comments which helped to improve the paper.
This work has been supported by the Verbundforschung (DLR) of the BMBF under Grant No. 50 OR 0203 and the Deutsche Forschungsgemeinschaft (DFG) under RE 353/49-1.
\end{acknowledgements}

\bibliographystyle{aa}
\bibliography{5556}

\begin{thebibliography}{50}
\expandafter\ifx\csname natexlab\endcsname\relax\def\natexlab#1{#1}\fi

\bibitem[{{Agafonova} {et~al.}(2005){Agafonova}, {Centuri{\'o}n}, {Levshakov},
  \& {Molaro}}]{agafonovaetal2005}
{Agafonova}, I.~I., {Centuri{\'o}n}, M., {Levshakov}, S.~A., \& {Molaro}, P.
  2005, \aap, 441, 9

\bibitem[{{Aguirre} {et~al.}(2004){Aguirre}, {Schaye}, {Kim}, {Theuns},
  {Rauch}, \& {Sargent}}]{aguirreetal2004}
{Aguirre}, A., {Schaye}, J., {Kim}, T., {et~al.} 2004, \apj, 602, 38

\bibitem[{{Ballester} {et~al.}(2000){Ballester}, {Mondigliani}, {Boitquin}, S.,
  {Hanuschik}, {Kaufer}, \& {Wolf}}]{ballesteretal2000}
{Ballester}, P., {Mondigliani}, A., {Boitquin}, O., {et~al.} 2000, ESO
  Messenger, 101, 31

\bibitem[{{Bianchi} {et~al.}(2001){Bianchi}, {Cristiani}, \&
  {Kim}}]{bianchietal2001}
{Bianchi}, S., {Cristiani}, S., \& {Kim}, T.-S. 2001, \aap, 376, 1

\bibitem[{{Boksenberg} {et~al.}(2003){Boksenberg}, {Sargent}, \&
  {Rauch}}]{boksenbergetal2003}
{Boksenberg}, A., {Sargent}, W. L.~W., \& {Rauch}, M. 2003, astro-ph/0307557

\bibitem[{{Bolton} {et~al.}(2006){Bolton}, {Haehnelt}, {Viel}, \&
  {Carswell}}]{boltonetal2006}
{Bolton}, J.~S., {Haehnelt}, M.~G., {Viel}, M., \& {Carswell}, R.~F. 2006,
  \mnras, 366, 1378

\bibitem[{{Cardelli} {et~al.}(1989){Cardelli}, {Clayton}, \&
  {Mathis}}]{cardellietal1989}
{Cardelli}, J.~A., {Clayton}, G.~C., \& {Mathis}, J.~S. 1989, \apj, 345, 245

\bibitem[{{Croft}(2004)}]{croft2004}
{Croft}, R.~A.~C. 2004, \apj, 610, 642

\bibitem[{{Erb} {et~al.}(2003){Erb}, {Shapley}, {Steidel}, {Pettini},
  {Adelberger}, {Hunt}, {Moorwood}, \& {Cuby}}]{erbetal2003}
{Erb}, D.~K., {Shapley}, A.~E., {Steidel}, C.~C., {et~al.} 2003, \apj, 591, 101

\bibitem[{{Fardal} {et~al.}(1998){Fardal}, {Giroux}, \&
  {Shull}}]{fardaletal1998}
{Fardal}, M.~A., {Giroux}, M.~L., \& {Shull}, J.~M. 1998, \aj, 115, 2206

\bibitem[{{Fechner} {et~al.}(2004){Fechner}, {Baade}, \&
  {Reimers}}]{fechneretal2004}
{Fechner}, C., {Baade}, R., \& {Reimers}, D. 2004, \aap, 418, 857

\bibitem[{{Fechner} {et~al.}(2006{\natexlab{a}}){Fechner}, {Reimers}, {Kriss},
  {Baade}, {Blair}, {Giroux}, {Green}, {Moos}, {Morton}, {Scott}, {Shull},
  {Simcoe}, {Songaila}, \& {Zheng}}]{fechneretal2006b}
{Fechner}, C., {Reimers}, D., {Kriss}, G.~A., {et~al.} 2006{\natexlab{a}},
  \aap, 455, 91

\bibitem[{{Fechner} {et~al.}(2006{\natexlab{b}}){Fechner}, {Reimers},
  {Songaila}, {Simcoe}, {Rauch}, \& {Sargent}}]{fechneretal2006a}
{Fechner}, C., {Reimers}, D., {Songaila}, A., {et~al.} 2006{\natexlab{b}},
  \aap, 455, 73

\bibitem[{{Fox} {et~al.}(2005){Fox}, {Savage}, \& {Wakker}}]{foxetal2005}
{Fox}, A.~J., {Savage}, B.~D., \& {Wakker}, B.~P. 2005, \aj, 130, 2418

\bibitem[{{Haardt} \& {Madau}(1996)}]{haardtmadau1996}
{Haardt}, F. \& {Madau}, P. 1996, \apj, 461, 20

\bibitem[{{Haardt} \& {Madau}(2001)}]{haardtmadau2001}
{Haardt}, F. \& {Madau}, P. 2001, in Clusters of Galaxies and the High Redshift
  Universe Observed in X-rays, ed. D.~M. {Neumann} \& J.~T.~T. {Van}, 64

\bibitem[{{Hui} \& {Gnedin}(1997)}]{huignedin1997}
{Hui}, L. \& {Gnedin}, N.~Y. 1997, \mnras, 292, 27

\bibitem[{{Jakobsen} {et~al.}(2003){Jakobsen}, {Jansen}, {Wagner}, \&
  {Reimers}}]{jakobsenetal2003}
{Jakobsen}, P., {Jansen}, R.~A., {Wagner}, S., \& {Reimers}, D. 2003, \aap,
  397, 891

\bibitem[{{K{\" o}hler} {et~al.}(1996){K{\" o}hler}, {Reimers}, \&
  {Wamsteker}}]{koehleretal1996}
{K{\" o}hler}, S., {Reimers}, D., \& {Wamsteker}, W. 1996, \aap, 312, 33

\bibitem[{{Kim} {et~al.}(2001){Kim}, {Cristiani}, \& {D'Odorico}}]{kimetal2001}
{Kim}, T.-S., {Cristiani}, S., \& {D'Odorico}, S. 2001, \aap, 373, 757

\bibitem[{{Kim} {et~al.}(2002){Kim}, {Cristiani}, \& {D'Odorico}}]{kimetal2002}
{Kim}, T.-S., {Cristiani}, S., \& {D'Odorico}, S. 2002, \aap, 383, 747

\bibitem[{{Kirkman} \& {Tytler}(1997)}]{kirkmantytler1997}
{Kirkman}, D. \& {Tytler}, D. 1997, \apj, 484, 672

\bibitem[{{Kriss} {et~al.}(2001){Kriss}, {Shull}, {Oegerle}, {Zheng},
  {Davidsen}, {Songaila}, {Tumlinson}, {Cowie}, {Deharveng}, {Friedman},
  {Giroux}, {Green}, {Hutchings}, {Jenkins}, {Kruk}, {Moos}, {Morton},
  {Sembach}, \& {Tripp}}]{krissetal2001}
{Kriss}, G.~A., {Shull}, J.~M., {Oegerle}, W., {et~al.} 2001, Science, 293,
  1112

\bibitem[{{Lai} {et~al.}(2006){Lai}, {Lidz}, {Hernquist}, \&
  {Zaldarriaga}}]{laietal2006}
{Lai}, K., {Lidz}, A., {Hernquist}, L., \& {Zaldarriaga}, M. 2006, \apj, 644,
  61

\bibitem[{{Liu} {et~al.}(2006){Liu}, {Jamkhedkar}, {Zheng}, {Feng}, \&
  {Fang}}]{liuetal2006}
{Liu}, J., {Jamkhedkar}, P., {Zheng}, W., {Feng}, L.-L., \& {Fang}, L.-Z. 2006,
  \apj, 645, 861

\bibitem[{{Maselli} \& {Ferrara}(2005)}]{maselliferrara2005}
{Maselli}, A. \& {Ferrara}, A. 2005, \mnras, 364, 1429

\bibitem[{{Miralda-Escud{\' e}}(1993)}]{miraldaescude1993}
{Miralda-Escud{\' e}}, J. 1993, \mnras, 262, 273

\bibitem[{{Miralda-Escud{\' e}}(2005)}]{miraldaescude2005}
{Miralda-Escud{\' e}}, J. 2005, \apjl, 620, L91

\bibitem[{{Miralda-Escud{\' e}} {et~al.}(2000){Miralda-Escud{\' e}},
  {Haehnelt}, \& {Rees}}]{miraldaescudeetal2000}
{Miralda-Escud{\' e}}, J., {Haehnelt}, M., \& {Rees}, M.~J. 2000, \apj, 530, 1

\bibitem[{{M{\o}ller} \& {Jakobsen}(1990)}]{mollerjakobsen1990}
{M{\o}ller}, P. \& {Jakobsen}, P. 1990, \aap, 228, 299

\bibitem[{{Picard} \& {Jakobsen}(1993)}]{picardjakobsen1993}
{Picard}, A. \& {Jakobsen}, P. 1993, \aap, 276, 331

\bibitem[{{Reimers} {et~al.}(2006){Reimers}, {Fechner}, {Kriss}, {Shull},
  {Baade}, {Moos}, {Songaila}, \& {Simcoe}}]{reimersetal_fuse}
{Reimers}, D., {Fechner}, C., {Kriss}, G., {et~al.} 2006, in Astronomical
  Society of the Pacific Conference Series, ed. G.~{Sonneborn}, H.~W. {Moos},
  \& B.-G. {Andersson}, 41--+, astro-ph/0410588

\bibitem[{{Reimers} {et~al.}(2005){Reimers}, {Hagen}, {Schramm}, {Kriss}, \&
  {Shull}}]{reimersetal2005b}
{Reimers}, D., {Hagen}, H.-J., {Schramm}, J., {Kriss}, G.~A., \& {Shull}, J.~M.
  2005, \aap, 436, 465

\bibitem[{{Reimers} {et~al.}(1997){Reimers}, {K{\"o}hler}, {Wisotzki},
  {Groote}, {Rodriguez-Pascual}, \& {Wamsteker}}]{reimersetal1997}
{Reimers}, D., {K{\"o}hler}, S., {Wisotzki}, L., {et~al.} 1997, \aap, 327, 890

\bibitem[{{Ricotti} {et~al.}(2000){Ricotti}, {Gnedin}, \&
  {Shull}}]{ricottietal2000}
{Ricotti}, M., {Gnedin}, N.~Y., \& {Shull}, J.~M. 2000, \apj, 534, 41

\bibitem[{{Savage} \& {Sembach}(1991)}]{savagesembach1991}
{Savage}, B.~D. \& {Sembach}, K.~R. 1991, \apj, 379, 245

\bibitem[{{Schaye}(2001)}]{schaye2001}
{Schaye}, J. 2001, \apj, 559, 507

\bibitem[{{Schaye}(2004)}]{schaye2004}
{Schaye}, J. 2004, \apj, submitted, astro-ph/0409137

\bibitem[{{Schaye} {et~al.}(1999){Schaye}, {Theuns}, {Leonard}, \&
  {Efstathiou}}]{schayeetal1999}
{Schaye}, J., {Theuns}, T., {Leonard}, A., \& {Efstathiou}, G. 1999, \mnras,
  310, 57

\bibitem[{{Schaye} {et~al.}(2000){Schaye}, {Theuns}, {Rauch}, {Efstathiou}, \&
  {Sargent}}]{schayeetal2000}
{Schaye}, J., {Theuns}, T., {Rauch}, M., {Efstathiou}, G., \& {Sargent},
  W.~L.~W. 2000, \mnras, 318, 817

\bibitem[{{Schirber} {et~al.}(2004){Schirber}, {Miralda-Escud{\' e}}, \&
  {McDonald}}]{schirberetal2004}
{Schirber}, M., {Miralda-Escud{\' e}}, J., \& {McDonald}, P. 2004, \apj, 610,
  105

\bibitem[{{Schlegel} {et~al.}(1998){Schlegel}, {Finkbeiner}, \&
  {Davis}}]{schlegeletal1998}
{Schlegel}, D.~J., {Finkbeiner}, D.~P., \& {Davis}, M. 1998, \apj, 500, 525

\bibitem[{{Shapley} {et~al.}(2005){Shapley}, {Steidel}, {Erb}, {Reddy},
  {Adelberger}, {Pettini}, {Barmby}, \& {Huang}}]{shapleyetal2005}
{Shapley}, A.~E., {Steidel}, C.~C., {Erb}, D.~K., {et~al.} 2005, \apj, 626, 698

\bibitem[{{Shull} {et~al.}(2004){Shull}, {Tumlinson}, {Giroux}, {Kriss}, \&
  {Reimers}}]{shulletal2004}
{Shull}, J.~M., {Tumlinson}, J., {Giroux}, M.~L., {Kriss}, G.~A., \& {Reimers},
  D. 2004, \apj, 600, 570

\bibitem[{{Simcoe} {et~al.}(2002){Simcoe}, {Sargent}, \&
  {Rauch}}]{simcoeetal2002}
{Simcoe}, R.~A., {Sargent}, W.~L.~W., \& {Rauch}, M. 2002, \apj, 578, 737

\bibitem[{{Smette} {et~al.}(2002){Smette}, {Heap}, {Williger}, {Tripp},
  {Jenkins}, \& {Songaila}}]{smetteetal2002}
{Smette}, A., {Heap}, S.~R., {Williger}, G.~M., {et~al.} 2002, \apj, 564, 542

\bibitem[{{Telfer} {et~al.}(2002){Telfer}, {Zheng}, {Kriss}, \&
  {Davidsen}}]{telferetal2002}
{Telfer}, R.~C., {Zheng}, W., {Kriss}, G.~A., \& {Davidsen}, A.~F. 2002, \apj,
  565, 773

\bibitem[{{Teplitz} {et~al.}(1998){Teplitz}, {Malkan}, \&
  {McLean}}]{teplitzetal1998}
{Teplitz}, H.~I., {Malkan}, M., \& {McLean}, I.~S. 1998, \apj, 506, 519

\bibitem[{{Vogel} \& {Reimers}(1995)}]{vogelreimers1995}
{Vogel}, S. \& {Reimers}, D. 1995, \aap, 294, 377

\bibitem[{{Zheng} {et~al.}(2004){Zheng}, {Kriss}, {Deharveng}, {Dixon}, {Kruk},
  {Shull}, {Giroux}, {Morton}, {Williger}, {Friedman}, \&
  {Moos}}]{zhengetal2004}
{Zheng}, W., {Kriss}, G.~A., {Deharveng}, J.-M., {et~al.} 2004, \apj, 605, 631

\end{thebibliography}
 
\end{document}